\documentclass[12pt]{article}
\usepackage[a4paper,left=1in,top=1in,width=6.3in,height=9in]{geometry}
\usepackage{graphicx} % Required for inserting images
\usepackage{bm}
\usepackage{tabularx}
\usepackage{authblk}
\usepackage[dvipsnames]{xcolor}
\usepackage{amsmath}
\usepackage{amsthm}
\usepackage{pdflscape}
\usepackage{mathtools}
\usepackage{amssymb}
\usepackage[title]{appendix}
\usepackage{multirow}
\usepackage[sort]{natbib}
\usepackage{subcaption}
\newcommand{\mycomment}[1]{}

\numberwithin{equation}{section}

\usepackage[width=.75\textwidth]{caption} %% added 14 Feb 2022
\numberwithin{equation}{section}
\usepackage[utf8]{inputenc}
\usepackage{url}
\usepackage{hyperref}
\usepackage{amsthm}
\newtheorem{theorem}{Theorem}
\newtheorem{definition}[theorem]{Definition}
 
\newcolumntype{a}{>{\columncolor{red}}c}
\newcolumntype{g}{>{\columncolor{green}}c}

\newcommand{\nil}[1]{}

\newcommand{\dummy}[1]{}
\providecommand{\keywords}[1]{\noindent\textbf{\textit{Keywords}}: #1}
\title{The projected isotropic normal  distribution  with applications in neuroscience }
\author{Kanti V. Mardia$^1$ and Antonio Mauricio F.L. Miranda de S$\Acute{a}$$^2$ }
\affil{$^{1}$ Department of Statistics, University of Leeds, Woodhouse, Leeds LS2 9JT, UK; k.v.mardia@leeds.ac.uk} 
\affil{$^2$ Biomedical Engineering Program, Federal University of Rio de Janeiro, Brazil; amflms@peb.ufrj.br} 
\begin{document}
\date{} %Commneted out will give the date
\maketitle
\begin{abstract}
This paper is motivated by a cutting‑edge application in neuroscience: the analysis of electroencephalogram (EEG) signals recorded under flash stimulation. Under commonly used signal‑processing assumptions, only the phase angle of the EEG is required for the analysis of such applications. We demonstrate that these assumptions imply that the phase has a projected isotropic normal distribution. We revisit this distribution and derive several new properties, including closed‑form expressions for its trigonometric moments. We then examine the distribution of the mean resultant and its square — a statistic of central importance in phase‑based EEG studies. The distribution of the resultant is analytically intricate; to make it practically useful, we develop two approximations based on the well‑known resultant distribution for the von Mises distribution. We then study inference problems for this projected isotropic normal distribution. The method is illustrated with an application to EEG data from flash‑stimulation experiments.
\end{abstract}

\keywords{component synchrony measure, EEG, phase angle, resultant, score matching approximation, von Mises distribution
}
 
\section{Introduction }\label{Intro}
Our paper is motivated by a cutting‑edge application in biomedical signal analysis, which deals with analysing biomedical signals. These signals are collected from the body at the organ, cellular, or molecular level. There are different biomedical signals, including the electroencephalogram (EEG), which reflects electrical activity from the brain, and the electrocardiogram (ECG), which reflects electrical activity from the heart, among others (see, for example, \citet{Muth2004}, \citet{Subasi2019}, and \citet{Anand2022}). ECG traces are well-known (as used for the heart). \citet{mushtaq2024eeg}, in reviewing one hundred years of EEG research, emphasise the importance of methodological progress by stating that “Future work should aim to develop EEG pipelines that are transparent, reproducible and generalizable across populations and contexts.” This call for improved workflows implicitly includes the adoption of new statistical tools and analytical frameworks capable of handling the complexity and variability inherent in EEG data.

The EEG signals are, in general, univariate time series \citep{Muth1998}. The Fourier transform of such a series provides a quantitative description of its frequency content. For neurological data, examining spectral or frequency‑domain characteristics is often as important as studying time‑domain behaviour. For example, changes in electroencephalograms (EEGs) across different stages of sleep are more clearly captured through frequency‑domain features. Although our application focuses on EEG responses to flash stimulation (Section~\ref{Appl}), the same principles apply across a wide range of experimental paradigms.

{\bf  Model.} To formalise this framework, we treat EEG recordings as traces or univariate continuous‑time series $x(t)$, where $t \in \mathbb{R}^+$ (the non‑negative real line), typically regarded as defined on $0 \le t < \infty$.

Under some plausible assumptions, \citet{Mir2002} and \citet{Mir2003} have proposed that the discrete Fourier transform $Y_j$ of $x(t)$ at frequency $j$ can be modelled as
\begin{equation}\label{spect}
Y_j = \mu + e_{1j} + i\, e_{2j}, \quad j = 1,2,\ldots,
\end{equation}
where $i = \sqrt{-1}$, and $e_{1j}$ and $e_{2j}$ are independent error terms.
Further, it has been shown in various studies that, for EEG applications like ours, the \emph{phase} of the Fourier transform plays the main role rather than its magnitude \citep{Busch2009,BuschVanRullen2010,Mathewson2009,NotbohmHerrmann2016}. That is, only the angular component $\theta$ of $Y_j$ in (\ref{spect}) is of interest. Although the joint model (\ref{spect}) was proposed by \citet{Mir2002} and \citet{Mir2003}, it was not recognised that the resulting phase distribution belongs to the family of projected normal distributions (see, for example, \citet{mardiajupp2000}). In Section~\ref{Sec: PIN}, we derive this distribution for our setting and refer to it as the projected isotropic normal (PIN) distribution. In Section~\ref{Sec: Pop}, we obtain population moments of the PIN distribution and use these to construct two approximations by the von Mises distribution: the standard approximation and a new approximation based on the score‑matching method. 

The well‑known summary statistic “mean resultant” $\bar{R}$ in directional statistics is found to capture the behaviour of the phase of the Fourier transform (see Section~\ref{Sec: Rbar}). In the biomedical literature, its square ($\bar{R}^2$) is also used and termed the “Component Synchrony Measure (CSM)”. To perform statistical inference on CSM, we need the sampling distribution of the resultant length $R$ under the PIN distribution, which is considered in Section~\ref{Sec: MainDistR}. It is seen that the distribution is complicated, but it can be approximated using the distribution under the von Mises case. For small and large values of the concentration parameter of the PIN distribution, both approximations are very good, and even for mid‑range values the loss is minimal (see Sections~\ref{Sec: Rbar} and \ref{CSMvM}).

In Section~\ref{Sec: Est}, we consider estimation of the parameters, including the maximum likelihood method; the maximum likelihood estimates can be obtained only using numerical methods, and we provide some approximate closed‑form estimates, including those based on the score‑matching method. Under a deterministic response to stimulation, the phase would be identical across trials, and the noise component would have a uniform phase distribution. Thus, the phase itself becomes the crucial quantity for formulating the null hypothesis of no response. In Section~\ref{Tests}, we consider hypothesis testing under the PIN distribution and provide confidence intervals for CSM. In Section~\ref{Appl}, we then apply these results to biomedical EEG signals recorded under flash stimulation. Finally, Section~\ref{Disc} highlights the key contributions of the paper and gives directions for further work. Some supplementary material has been  given in Appendices ; Appendix 1  \ref{Sec: Appendix1Model} examines the EEG model in detail;, Appendix 2 \ref{Sec: Appendix2Moment} gives the proof of moments formula for the PIN distribution and Appendix 3 \ref{Sec: Appendix3Approx12} examines  further the two  approximations to the PIN distribution by  the von Mises distribution.

\section{ The Projected Isotropic Normal  (PIN) distribution}\label{Sec: PIN}

Let $(x_1, x_2)^T$ be distributed as an isotropic bivariate normal distribution with mean vector $(\mu_1, \mu_2)^T$ and covariance matrix $\sigma^2 \bm{I}$  where $\bm{I}$ is the identity matrix. Let
\begin{equation}\label{polar}
    x_1 = r \cos \theta,\quad x_2 = r \sin \theta,\quad r > 0,\; 0 \le \theta \le 2\pi,
\end{equation}
then $\theta$, in terms of the Fourier transform model, is the \emph{phase angle}. This distribution of $\theta$ is a particular case of the general projected normal distribution given in \citet{mardia1972}, page~92, equation~(3.4.17), under a bivariate normal with general mean vector and general covariance matrix. To deduce the projected normal distribution for our case, in that equation we substitute
\begin{equation}
    \mu = \mu_1 = \beta \cos \mu,\quad \nu = \mu_2 = \beta \sin \mu,\quad
    \sigma_1^2 = \sigma_2^2 = \sigma^2,\quad \rho = 0,
\end{equation}
which leads to the following form of the projected normal distribution, where $\gamma = \beta^2/(4\sigma^2)$.

\begin{definition}
The \emph{Projected Isotropic Normal} (PIN) distribution of $\theta$ has the following probality density function (pdf):
\begin{equation}\label{PIN}
    f(\theta; \mu, \gamma) =
    \frac{1}{2\pi} \exp(-2\gamma)
    + 2\sqrt{\gamma}\cos(\theta - \mu)\,\Phi(2\sqrt{\gamma}\cos(\theta - \mu))\,
    \phi(2\sqrt{\gamma}\sin(\theta - \mu)),
\end{equation}
where $-\pi \le \theta \le \pi$, $\gamma \ge 0$, and $\gamma$ is the “concentration parameter’’ of the PIN distribution. We say that $\theta$ is distributed as PIN($\mu,\gamma$), or simply PIN($\gamma$) if $\mu = 0$.
\end{definition}

We sometimes write
\begin{equation}\label{Eq: SNR}
 \text{SNR} = 2\gamma, \gamma = \tfrac{1}{2}\,\text{SNR},   
\end{equation}

where SNR stands for “Signal‑to‑Noise Ratio”, defined already in Section~\ref{Intro}.

We note that projected normal distributions are well established (see, for example, \citet{mardiajupp2000}). The distribution for the circular case in general was given by \citet{mardia1972}, with some particular cases. Note that for the circular case it is also called the off‑set normal distribution by \citet{mardiajupp2000}, the displaced normal by \citet{Kendall1974}, and the angular normal by \citet{Watson1983}.

 In practice, the von Mises distribution, the wrapped normal distribution, and the PIN distribution can each qualify as a “circular normal”. Each has its own merits; for example, the PIN and wrapped normal distributions are the easiest to simulate, while for testing uniformity the von Mises distribution leads to the Rayleigh test. Note that all three can be derived from the bivariate normal or the univariate normal: the von Mises distribution is the conditional distribution of $\theta$ given $r$ in $(r,\theta)$, the PIN distribution is the marginal distribution of $\theta$ in $(r,\theta)$, and the wrapped normal distribution is obtained by wrapping the univariate normal distribution.

It is not surprising that all three have been compared using the moment estimator of the concentration parameter (with $\mu = 0$ without any loss of generality), starting from \citet{Kendall1974}. \citet{Watson1983} (and recently summarised by \citet{Presnell1998}) made a thorough comparison and concluded that the maximum difference occurs at $\rho = .785\; (\kappa = 2.7,\; \gamma = 0.64)$, where $\rho$ denotes the population resultant. Section~\ref{Sec: VmApprox} studies further the PIN distribution approximated by a von Mises distribution.

We now give some properties of the PIN distribution. As $\gamma \to \infty$, $\Phi(2\sqrt{\gamma}\cos \theta) \to 1$, so from (\ref{PIN}) we obtain
\begin{equation}
    f(\theta;\gamma) \propto \sqrt{\gamma}\,|\cos \theta|\,\exp(-2\gamma \sin^2 \theta)\,d\theta,
\end{equation}
that is,
\begin{equation}
    f(\theta) \propto \exp(-2\gamma \theta^2),
\end{equation}
so $\theta$ is distributed as $N(0, 1/(4\gamma))$, which would mean that $4\gamma = 2\,\text{SNR}$ is the concentration parameter, although we take $\gamma$ as our concentration parameter. It can be seen that
\[
f(-\theta;\gamma) = f(\theta;\gamma),
\]
so the distribution is symmetrical about $\theta = 0$. Further, let $2\sqrt{\gamma} = \delta$. Then the derivative of $f(\theta;\gamma)$ with respect to $\theta$ is
\[
-2\delta \sin \theta \left[(1+\delta)\,\Phi(\delta\cos \theta)\,\phi(\delta \sin \theta)
+ \delta^2 \cos^2 \theta\,\Phi(\delta\cos \theta)\right].
\]
Since the terms in square brackets are positive, the mode can be verified to be at $\theta = 0$. The value of the pdf at the mode is
\[
\frac{1}{2\pi}\exp(-2\gamma) + \sqrt{2\gamma/\pi}\,\Phi(2\sqrt{\gamma}),
\]
and the antimode at $\theta = \pi$ is
\[
\frac{1}{2\pi}\exp(-2\gamma) - \sqrt{2\gamma/\pi}\,\Phi(2\sqrt{\gamma}).
\]

\textbf{Simulation.} It is straightforward to simulate a sample from the PIN distribution. Most statistical software includes a routine to simulate a normal distribution, so we can draw a sample from
\[
x_1 \sim N(\mu_1, \sigma^2), \qquad x_2 \sim N(\mu_2, \sigma^2),
\]
and obtain the sample value of $\theta$ from (\ref{polar}). We note that, since we are starting from a bivariate normal distribution, a simple method to simulate is via the well‑known Box–Muller transformation.

\section{Population Moments and the von Mises Approximations}\label{Sec: Pop}

We now obtain the trigonometric moments of the PIN distribution, which will be used below to obtain approximations of the PIN distribution by a von Mises distribution. We assume, without any loss of generality, that the mean direction is zero for both distributions. It can be shown, after some tedious algebra, that the following hold:
\begin{equation}\label{moment1}
E(\sin\theta)=0,\quad E(\sin 2\theta)=0,\quad
E(\cos\theta)=\sqrt{\pi \gamma/2}\exp(-\gamma)\{ I_0(\gamma)+I_1(\gamma) \},
\end{equation}
where $I_p(\cdot)$ is the modified Bessel function of the first kind and of order $p$, given by
\begin{equation}
    I_p(z)= \sum_{m=0}^{\infty} \frac{\left(\frac{1}{4}z^2 \right)^{m+\frac{p}{2}}}{m!\,\Gamma(p+m+1)}.
    \label{eq:BesselI}
\end{equation}
This result was first stated in \citet{Kendall1974}. Further, the population resultant $\rho$ is simply $E(\cos\theta)$; note that here $E(\cos\theta)\ge 0$. We also have $E(\cos\theta \sin\theta)=\tfrac{1}{2}E(\sin 2\theta)=0$ (see below).

In fact, we  extend the above results to higher moments, and the results are as follows.
\begin{theorem}
\begin{align}
E(\sin p\theta) &= 0, \label{eq:sinMoment}\\
E(\cos p\theta) &= \sqrt{\pi \gamma/2}\exp(-\gamma)\{ I_{(p-1)/2}(\gamma) + I_{(p+1)/2}(\gamma) \}. \label{eq:cosMoment}
\end{align}
\end{theorem}
where $p$ is an integer.

A proof is given in Appendix~2.

Note that for even moments, (\ref{eq:cosMoment}) can be simplified as follows. With $p=2k$, using \citet{abramowitz1964}, Equation (10.2.4), p.~445, we have
\[
\sqrt{\pi \gamma/2}\,I_{k+1/2}(\gamma)
= \gamma^k \left(\frac{1}{\gamma}\frac{d}{d\gamma}\right)^k \frac{\sinh \gamma}{\gamma}.
\]
Using this expression for $k=0,1$ in (\ref{eq:cosMoment}) with $p=2$, it can be shown that
\begin{equation}\label{Cos2}
    E(\cos 2\theta)= 1 - \frac{\exp(-\gamma)}{\gamma}\sinh \gamma.
\end{equation}
It is interesting that, in this case, this second cosine moment is simpler than the first cosine moment given by (\ref{moment1}).

\subsection{The von Mises Approximations}\label{Sec: VmApprox}
On the circle, the von Mises distribution (like the normal distribution) has two parameters, with the pdf of the angle $\theta$ given by
\begin{equation}\label{vM}
    f(\theta;\mu,\kappa)=\frac{\exp\{\kappa \cos(\theta-\mu)\}}{2\pi I_0(\kappa)},\quad 0\le \theta< 2\pi,
\end{equation}
where $\mu$ is the mean direction, $0\le \mu<2\pi$, and $\kappa$ is the concentration (“inverse variance”) parameter, $\kappa\ge 0$, with $I_0(\kappa)$ the modified Bessel function of the first kind and order zero given in (\ref{eq:BesselI}). This distribution is symmetrical with mode at $\mu$. For $\kappa=0$, $\theta$ is uniformly distributed on the circle. For large $\kappa$, $\theta$ is approximately normal with mean $\mu$ and variance $2/\kappa$. We denote this distribution by vM($\mu,\kappa$), and when $\mu = 0$ we denote the distribution by vM($\kappa$).

We now show how to approximate the PIN distribution PIN($\gamma$) by the von Mises distribution vM($\kappa$), with concentration parameter $\kappa$.

\subsubsection{The Standard von Mises Approximation: Approx 1}\label{Sec: VmApprox 1}

The standard approximation consists of equating the moment $E(\cos\theta)$ of the two distributions. That is,
\begin{equation}\label{Approx 1}
  A(\kappa) = \sqrt{\pi \gamma/2}\exp(-\gamma)\{ I_0(\gamma)+I_1(\gamma) \}
  \quad\text{or}\quad
  \kappa = A^{-1}\!\left(\sqrt{\pi \gamma/2}\exp(-\gamma)\{ I_0(\gamma)+I_1(\gamma) \}\right),
\end{equation}
where $A(\kappa)=I_1(\kappa)/I_0(\kappa)$ and $I_p(\cdot)$ denotes the modified Bessel function of the first kind and order $p$ given in (\ref{eq:BesselI}). We refer to the approximation defined by (\ref{Approx 1}) as Approx~1.

From (\ref{Approx 1}), it is shown below that
\begin{equation}\label{small1}
  \text{for small }\kappa\ \text{or for small }\gamma,\qquad \kappa=\sqrt{2\pi \gamma},
\end{equation}
whereas
\begin{equation}\label{large1}
  \text{for large }\kappa\ \text{or for large }\gamma,\qquad \kappa = 4\gamma.
\end{equation}
To derive these  small‑ and large‑$\kappa$ expressions for Approx~1, we use standard expansions for the modified Bessel functions of the first kind, as given in Appendix~1 of \citet{mardiajupp2000}.
 We substitute into (\ref{Approx 1}) the following approximations and the proof follows:
\begin{equation}\label{Bsmall}
\text{For small }\kappa:\qquad
I_0(\kappa) \approx 1,\qquad
I_1(\kappa) \approx \kappa.
\end{equation}
\begin{equation}\label{Blarge}
\text{For large }\kappa:\qquad
I_0(\kappa)
\approx \frac{\exp(\kappa)}{\sqrt{2\pi\kappa}}
\left(1 + \frac{1}{8\kappa}\right),\qquad
I_1(\kappa)
\approx \frac{\exp(\kappa)}{\sqrt{2\pi\kappa}}
\left(1 - \frac{3}{8\kappa}\right).
\end{equation}

\begin{figure}[!htb]
\begin{center}
\includegraphics[width=.9\textwidth]{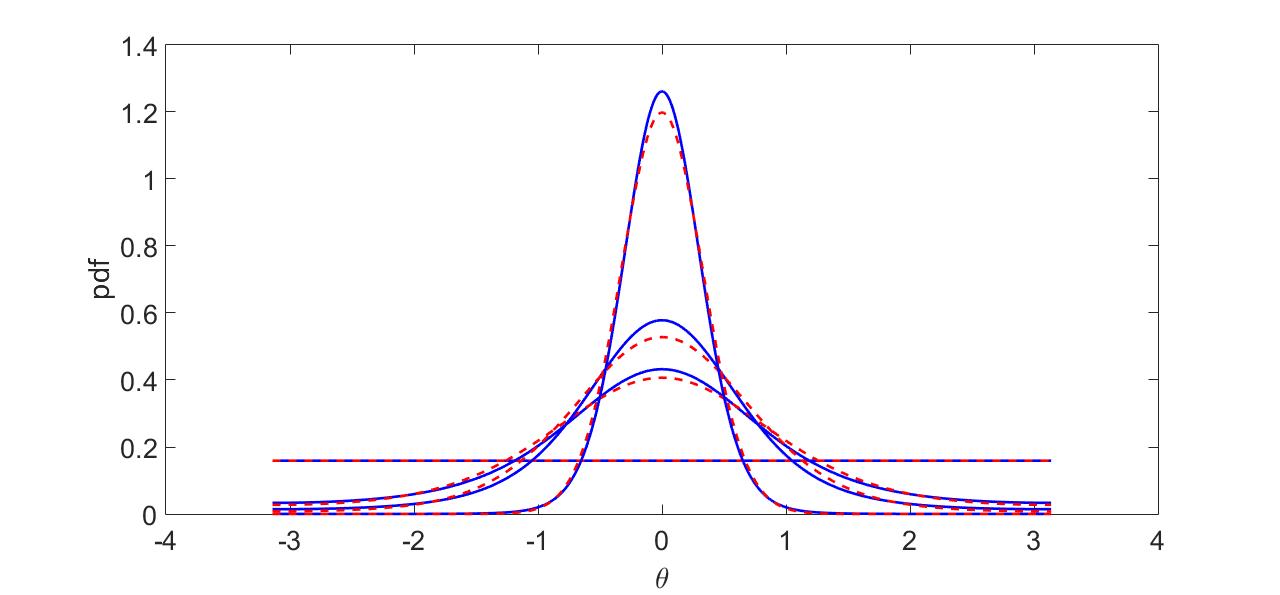}
\end{center}
\caption{\label{theta-PINnew}.  Projected isotropic normal (PIN) distribution for different values of $\gamma$ ($0, 0.25, 0.5, 2.5$) shown in continuous blue lines, and the corresponding von Mises distribution approximation (Approx~1) shown in dashed red lines. }

\end{figure}

\subsubsection{Score Matching von Mises Approximation: Approx 2}\label{Sec: VmApprox2}
We now derive the Score Matching Approximation (SMA) as developed in \citet{Mardia2017} and \citet{Mardia2018}. From \citet{Mardia2017},
\begin{equation}\label{SMA}
\kappa = \frac{2E(\cos\theta)}{1 - E(\cos 2\theta)} .
\end{equation}
In \citet{Mardia2018}, the same equation (equation (19)) contains a typo: in the right-hand side of this equation, the factor 2 should be deleted, and the same correction applies to the particular cases given in (20) and (21).

On substituting $E(\cos\theta)$ from (\ref{moment1}) and $E(\cos 2\theta)$ from (\ref{Cos2}) into (\ref{SMA}), we obtain our second approximation:
\begin{equation}\label{Approx 2}
\kappa = \frac{\gamma\sqrt{2\pi \gamma}\,\{ I_0(\gamma)+I_1(\gamma) \}}{\sinh \gamma}.
\end{equation}
 We refer to this approximation defined by (\ref{Approx 2}) as Approx~2.
From (\ref{Approx 2}), we show that
\begin{equation}\label{small2}
\text{for small }\kappa\ \text{or for small }\gamma,\qquad
\kappa = \sqrt{2\pi \gamma},
\end{equation}
and further,
\begin{equation}\label{large2}
\text{for large }\kappa\ \text{or for large }\gamma,\qquad \kappa = 4\gamma.
\end{equation}
To derive thees small‑ and large‑$\kappa$ expressions we again  use some standard expansions for the modified Bessel functions of the first kind, as given in Appendix~1 of \citet{mardiajupp2000}. We substitute into (\ref{Approx 2}):

- For small $\kappa$, use (\ref{Bsmall}) and $(\sinh\gamma)/\gamma \approx 1$.  
- For large $\kappa$, use (\ref{Blarge}) and $\sinh\gamma \approx \exp(\gamma)/2$.

Comparing (\ref{small1})–(\ref{large1}) with (\ref{small2})–(\ref{large2}) shows that Approx~1 and Approx~2 are identical in both the small‑ and large‑$\kappa$ cases.

We note that Approx~2 is simpler than Approx~1, since in Approx~2 the left-hand side of equation (\ref{Approx 2}) does not involve any Bessel function (cf. (\ref{Approx 1})). However, we have found that their overall performance is similar when applied to problems involving the PIN distribution; these details are given in Appendix 3 \ref{Sec: Appendix3Approx12} . Mostly, we will concentrate on Approx~1, except when we need the computational advantage of Approx~2. 

\section{Distribution of the Mean Resultant}\label{Sec: Rbar}

Let $\theta_1, \ldots, \theta_n$ be a random sample drawn from PIN($\gamma$), and let us write
\begin{equation}\label{CS}
  \bar{C} = \frac{1}{n}\sum_{i=1}^{n} \cos \theta_i,\qquad
  \bar{S} = \frac{1}{n}\sum_{i=1}^{n} \sin \theta_i,
\end{equation}
Then the sample mean resultant $\bar{R}$ is given by
\begin{equation}\label{R}
  \bar{R} = \sqrt{\bar{C}^2 + \bar{S}^2}.
\end{equation}

We now give the different nomenclature used in the EEG literature for the mean resultant, to highlight its target application. We start with the term used for $\bar{R}^2$.
\begin{definition}
The quantity $\bar{R}^2$ is the “component synchrony measure’’ (CSM) of the output signal, denoted by $\hat{\rho}^2(f)$. CSM is also known as “phase synchrony’’ or “phase synchrony measure’’; see, for example, \citet{Fried1984}.
\end{definition}
That is,
\begin{equation}\label{CSF}
\text{CSM} = \bar{R}^2 = \bar{C}^2 + \bar{S}^2 = R^2/n^2 = \hat{\rho}^2(f),
\end{equation}
where $f$ in $\hat{\rho}^2(f)$ highlights that these quantities arise from the frequencies of the EEG trace. In psychology, synchrony refers to the spontaneous rhythmic coordination of actions, emotions, thoughts, and physiological processes across time between two or more individuals.

Furthermore, the mean resultant appears with terminology specific to EEG data in \citet{cohen2014EEG} as follows: the Phase Locking Value (PLV) is the name used for the mean resultant. In particular, the term Inter-Site Phase Clustering (ISPC) (\citet{cohen2014EEG}, Chapter~19) is used for PLV computed for a single electrode site, whereas the same term is also used for PLV computed from phase-angle differences for two electrode sites. “Clustering’’ here is synonymous with concentration.

Our aim is to obtain the distribution of $\bar{R}$ under the PIN distribution. We start with asymptotics, recalling that it is uniform when $\gamma = 0$.

\subsection{Asymptotics}\label{Sec: AsmRDist}
It is well-known that under uniformity, for large $n$, we have (see, for example, \citet{mardiajupp2000}, page~77)
\begin{equation}\label{Unif}
  2n\bar{R}^2 \asymp \chi^2_2.
\end{equation}
This result depends on the fact that $(\bar{C}, \bar{S})$ are asymptotically independent normals with means $0$ and variances $1/(2n)$. Note that in this uniform case, the exact pdf can be written down; see Section~\ref{Sec: MainDistR} below.
In general, we have from \citet{mardiajupp2000}, page~76, the asymptotic distribution of $\bar{R}$ as a normal distribution. In fact, from page~76 of \citet{mardiajupp2000}, we find that in our case, for large $n$, $(\bar{C}, \bar{S})$ are asymptotically independent normals with means and variances as follows:
\begin{equation}\label{CSlarge}
  E(\bar{C}) = \alpha,\qquad E(\bar{S}) = 0,\qquad
  \mathrm{var}(\bar{C}) = \frac{1}{2n}(1+\alpha_2),\qquad
  \mathrm{var}(\bar{S}) = \frac{1}{2n}(1-\alpha_2),
\end{equation}
where $\alpha = E(\cos\theta)$ and $\alpha_2 = E(\cos 2\theta)$.

Note that using (\ref{Unif}), it can be shown that for large $n$ we have
\begin{equation}\label{Resultantlarge}
E(\bar{R}) = \sqrt{\frac{\pi}{4n}},\qquad
E(\bar{R}^2) = \frac{1}{n},\qquad
\mathrm{var}(\bar{R}) = \frac{1}{n}\left(1 - \frac{\pi}{4}\right),\qquad
\mathrm{var}(\bar{R}^2) = \frac{1}{n^2}.
\end{equation}
\subsection{Distributions of the Resultant}\label{Sec: MainDistR}
Let $R = n\bar{R}$. The pdf  of $R$ under \textbf{uniformity} is (see \citet{mardiajupp2000}, page~66)
\begin{equation}\label{UniExact}
h_n(R) = R \int_0^{\infty} u\, J_0(Ru)\, J_0(u)^n\, du,
\end{equation}
where $J_0(\cdot)$ denotes the Bessel function of the first kind and order zero.
Also, we note that the pdf of $R$ for \textbf{the von Mises case} with concentration parameter $\kappa$ is given by \\(see \citet{mardiajupp2000}, page~69)
\begin{equation}\label{VMExact}
p(R) = \frac{I_0(\kappa R)}{I_0(\kappa)^n}\, h_n(R),\qquad 0 < R < n,
\end{equation}
where $h_n(R)$ is given by (\ref{UniExact}). Hence, using the simple transformation (\ref{CSF}) from $R$ to $v = \bar{R}^2 = R^2/n^2$ ($v = \mathrm{CSM} = \hat{\rho}^2(f)$), the distribution of $v$ under the von Mises distribution is
\begin{equation}\label{VMExactCSM}
p(v) = \frac{n^2}{2}\,
\frac{I_0(n\kappa\sqrt{v})}{I_0(\kappa)^n}
\int_0^{\infty} u\, J_0(n\sqrt{v}\,u)\, J_0(u)^n\, du,
\qquad 0 < v < 1.
\end{equation}
Note that for $\kappa \ge 4$, a very good approximation to the distribution of $R$ for the von Mises case (\ref{VMExactCSM}) is given by \citet{stephens1969tests}:
\begin{equation}\label{VMApproxCSM}
2n\gamma^*(1 - \bar{R}) \sim \chi^2_{n-1},
\end{equation}
where
\begin{equation}\label{ApprGamma}
\frac{1}{\gamma^*} = \frac{1}{\kappa} + \frac{3}{8\kappa^2}.
\end{equation}
Further, the percentage points of this distribution for the von Mises case are given in \citet{stephens1969tests}.
\section{The CSM Distribution and von Mises Approximations}\label{CSMvM}
In the last section, we gave an explicit expression for the sampling distribution of $R$ (or CSM) under the von Mises population, but the exact sampling distribution under the PIN distribution is not tractable, as we indicate even for the simplest case $n=2$.
The distribution of the resultant length $R$ from \citet{KentMardiaRao1979} for $n=2$, with a sample from a population with pdf $f(\cdot)$ (the CSM is a function of $R$), is given from Equation (2.4) as
\begin{equation}\label{Rforn=2}
\frac{4\int_0^{2\pi} f(u-t)f(u+t)\,du}{(4-R^2)^{1/2}},\qquad 0 < R < 2,
\end{equation}
where $t = \arccos(R/2)$. It can be shown that for the von Mises case, this expression simplifies to (\ref{VMExact}) for $n=2$, noting that
\begin{equation}\label{Eq: Uniformn=2}
h_2(R) = \frac{2}{\pi(4-R^2)^{1/2}},\qquad 0 < R < 2.
\end{equation}
However, even for this simplest case, (\ref{Rforn=2}) cannot be simplified for the PIN distribution.
As a practical solution, we now assume that Approx~1 provides the pdf of CSM for given values of $\gamma$ under the PIN distribution by plugging in the value of $\kappa$ corresponding to $\gamma$ from Approx~1 into (\ref{VMExactCSM}) for $\gamma > 0$,

\begin{equation}\label{VMapproxCSM}
p(v;\gamma) = \frac{n^2}{2}\,
\frac{I_0(n\kappa(\gamma)\sqrt{v})}{I_0(\kappa(\gamma))^n}
\int_0^{\infty} u\, J_0(nu\sqrt{v})\, J_0(u)^n\, du,
\qquad 0 < v < 1,\ \gamma > 0,
\end{equation}
where
\[
\kappa(\gamma) = A^{-1}\!\left(\sqrt{\pi\gamma/2}\exp(-\gamma)\{I_0(\gamma)+I_1(\gamma)\}\right),
\]
and  using (\ref{Unif}) for $\gamma = 0$
\begin{equation}\label{VMapproxCSmgamma0}
p(v;\gamma=0) = 2\int_0^{\infty} u\, J_0(u\sqrt{v})\, J_0(u)^n\, du.
\end{equation}

\begin{figure}[!htb]
\begin{center}
\includegraphics[width=.9\textwidth]{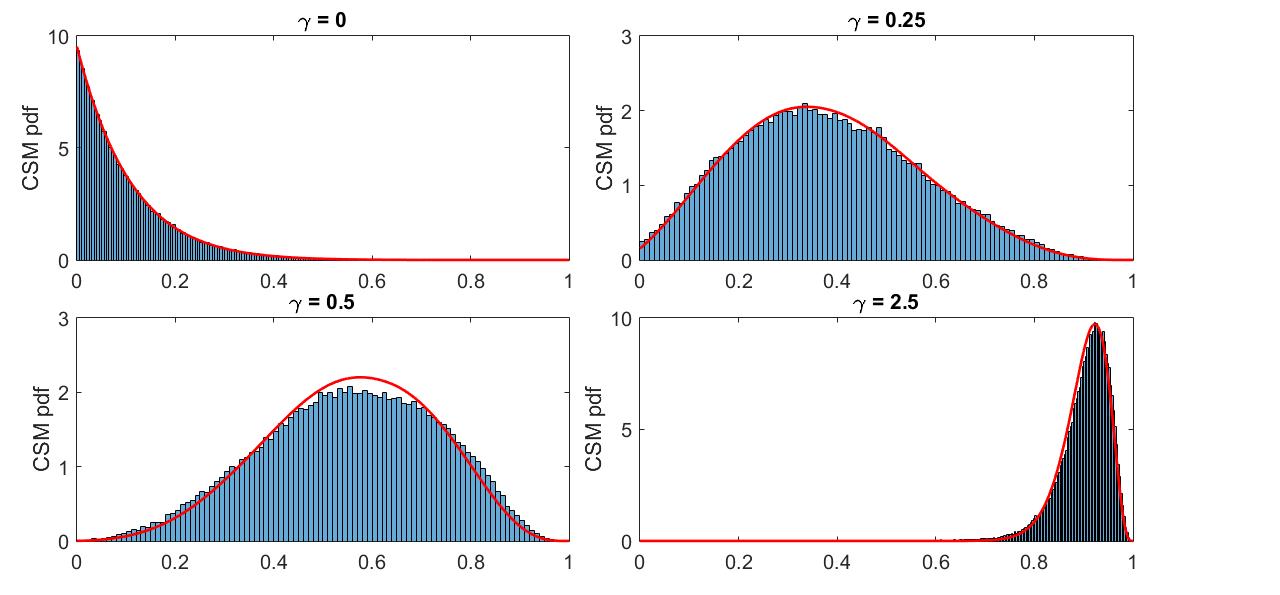}
\end{center}
\caption{\label{HistNew}
Simulated CSM histograms for $\gamma = 0, 0.25, 0.5, 2.5$ from the PIN density (blue, shaded), with the density (\ref{VMapproxCSM}) of CSM (red solid outline) obtained using Approx~1.
}
\end{figure}

To assess this approximation, we simulated the distribution of CSM (or $R$) from the distribution of $\theta$ under the PIN distribution. Simulation of $\theta$ is straightforward: given $\gamma$, draw two independent normal variables
\[
x = r\cos\theta \sim N(\sqrt{4\gamma}, 1),\qquad
y = r\sin\theta \sim N(0, 1),
\]
and compute $\theta = \arctan2(y, x)$ (the function $\arctan2$ returns the correct quadrant, so $0 \le \theta \le 2\pi$).
We used $n=10$ and $100{,}000$ simulations for
$\gamma = 0,\ 0.25,\ 0.5,\ 2.5,$
with the corresponding $\kappa$ values from Approx~1: $(0,\ 1.4,\ 2.1,\ 9.3)$.
Figure~\ref{HistNew} shows the simulated CSM histograms from the PIN density (blue, shaded), together with the approximate CSM density (red, solid outline)  provided by Approx~1, equation (4.9). It can be seen that the approximation is very good for small and large values of $\gamma$, and acceptable for the mid‑range. Note that we selected $n=10$ as such low values are common in EEG stimulation studies (see also Section \ref{Appl} with $n=12$.)
{\bf Remark}
  In EGG, if we allow the time dependent between observations, the distribution of the resultant length $R$ even for n=2 will be influenced by dependence. Let us assume for the angles on torus $(\theta_1, \theta_2) $ have the Mardia's cosine distribution (\cite{mardia1975},\cite{mardia1975b}) 
\begin{equation}\label{Eq: coorRuniform}
    f (\theta_1, \theta_2)= \frac{1}{(2\pi)^2 I_0(\lambda)} \exp{ \{ \lambda\cos (\theta_1-\theta_2)\} }
\end{equation}
where the association parameter  $\lambda >0$ favours $\theta_1 \approx \theta_2$ and hence larger $R$, while  $\lambda<0$ favours $\theta_1 \approx \theta_2 + \pi$ and shifts mass towards smaller $R$.
Note that the resultant for n=2  of  the two angles $\theta_1$ and $\theta_2$ can be seen to 
$$R=2|\cos(\theta_1-\theta_2)/2|.$$

  Now 
from \citet{KentMardiaRao1979} for $n=2$, with a sample from a population with pdf $f(\cdot)$ (the CSM is a function of $R$), can be extended  from Equation (2.4) to  as
\begin{equation}\label{Rforn=2}
\frac{4\int_0^{2\pi} f(u-t, u+t)\,du}{(4-R^2)^{1/2}},\qquad 0 < R < 2,
\end{equation}
where $t = \arccos(R/2)$.
It can be shown that for the cosine von Mises case, this cosine expression simplifies   for $n=2$ to, 

\begin{equation}\label{Eq: coorRuniform}
  f(R)  = \frac{2I_0(2\lambda \arccos(R/2))}{\pi I_0(\lambda)(4-R^2)^{1/2} }, \qquad 0 < R < 2.
\end{equation}
For $\lambda \to 0$, it reduces to the uniform distribution case given at (\ref{Eq: Uniformn=2}).

\section{Estimation}\label{Sec: Est}
Let $\theta_i,\ i=1,\ldots,n$ be a random sample from the PIN distribution. Then the maximum likelihood estimates (mles) $\hat{\mu},\hat{\gamma}$ of $\mu,\gamma$ are obtained by solving the (log-)likelihood equation from (\ref{PIN}):
\begin{equation}\label{LLM}
 \log L(\gamma) = \sum \log f(\theta_i;\mu,\gamma).
\end{equation}
However, this expression cannot be simplified so the mles $\hat{\mu},\hat{\gamma}$ must be computed numerically. We now give some approximation procedures.

\textbf{Hybrid Estimates.} Let
\begin{equation}\label{Mean and R}
\bar{C} = \bar{R}\cos\bar{\theta},\qquad
\bar{S} = \bar{R}\sin\bar{\theta},
\end{equation}
where $\bar{\theta}$ is the sample mean direction. One procedure is to plug into (\ref{LLM})
\begin{equation}\label{Mean}
\mu = \bar{\theta},
\end{equation}
and then maximise the function with respect to $\gamma$, leading to the hybrid estimate $\hat{\gamma}_{\text{HYB}}$.

\textbf{Moment Estimators.}
Let us write
\begin{equation}\label{rho}
E(\sin\theta) = \rho \sin\mu,\qquad
E(\cos\theta) = \rho \cos\mu,
\end{equation}
where $\rho$ is now a function of $\gamma$, written as $\rho(\gamma)$.
For $\mu$, we take the sample mean direction as a moment estimator:
\[
\hat{\mu}_{\text{MOM}} = \bar{\theta}.
\]
The simplest moment estimator of $\rho$ is
\begin{equation}\label{rhoMom}
\hat{\rho}_{\text{MOM}} = \bar{R},
\end{equation}
which leads to a moment estimator $\hat{\gamma}_{\text{MOM}}$ of $\gamma$ by substituting the value of $\rho$ from (\ref{moment1}):
\begin{equation}\label{vmEst}
\bar{R} = \sqrt{\pi\gamma/2}\exp(-\gamma)\{I_0(\gamma) + I_1(\gamma)\}.
\end{equation}
It can be shown  that
\begin{equation}\label{small}
\text{for small }\gamma,\qquad \hat{\gamma}_{\text{MOM}} = \frac{2\bar{R}^2}{\pi},
\end{equation}
whereas
\begin{equation}\label{large}
\text{for large }\gamma,\qquad \hat{\gamma}_{\text{MOM}} = \frac{1}{8(1-\bar{R})}.
\end{equation}
Approx~2 provides another “moment’’ estimate of $\gamma$ by using, for the von Mises distribution, $A(\hat{\kappa}) = \bar{R}$ in Approx~2. Thus $\hat{\gamma}_{\text{MOM2}}$ is the solution of
\begin{equation}\label{vmEst2}
\bar{R} = A\!\left(\frac{\sqrt{2\pi\gamma}\,\{I_0(\gamma)+I_1(\gamma)\}}{(\sinh\gamma)/\gamma}\right).
\end{equation}

\textbf{MLE of CSM.} In some applications, we need to obtain the mle of the population CSM $\rho^2$. Using the expression for $\rho$ from (\ref{moment1}), the mle $\hat{\rho}_{\text{MLE}}^2$ of $\rho^2$ is
\begin{equation}\label{rhoMLE}
\hat{\rho}_{\text{MLE}}^2
= (\pi \hat{\gamma}/2)\exp(-2\hat{\gamma})\{I_0(\hat{\gamma}) + I_1(\hat{\gamma})\}^2.
\end{equation}

\textbf{Bias.} We can study the bias in the approximate mles of $\gamma$ via well‑known results for the von Mises case related to $\kappa$. The estimator $\hat{\kappa} = A^{-1}(\bar{R})$ can be heavily biased for $n \le 15$, and a modified estimator $\hat{\kappa}^*$ is given by (see, for example, \citet{fisher1993}, \citet{mardiajupp2000})
\begin{equation}\label{UnbiasVMest}
\hat{\kappa}^* =
\begin{cases}
\max\{\,\hat{\kappa} - 2(n\hat{\kappa})^{-1},\, 0\,\}, & \hat{\kappa} < 2,\\[6pt]
\dfrac{(n-1)^3 \hat{\kappa}}{n^3 + n}, & \text{otherwise}.
\end{cases}
\end{equation}

\textbf{Asymptotic Distribution of $\hat{\gamma}$.}
Since our main focus is on the concentration parameter $\gamma$ rather than $\mu$, we consider only the asymptotic distribution of its estimator $\hat{\gamma}$. For large $n$,
\[
\hat{\gamma} \sim N(\gamma,\ 1/(nI)),
\]
where $I$ is the Fisher information,
\[
I = E\!\left[\frac{\partial^2}{\partial \gamma^2}\log L(\gamma)\right],
\]
and $\log L(\gamma)$ is given by (\ref{LLM}). There is no analytical simplification of $I$, so we resort to approximations.
In fact, we can show that
\[
\hat{\gamma}_{\text{MOM}} \sim N\!\left(\gamma,\ \frac{1}{n\,\mathrm{var}(\hat{\gamma}_{\text{MOM}})}\right),
\]
where $\mathrm{var}(\hat{\gamma}_{\text{MOM}})$ is derived as follows. Let $\hat{\rho} = \bar{R}$. From a Taylor approximation,
\begin{equation}\label{TaylorVar}
\mathrm{var}(\hat{\gamma}_{\text{MOM}})
= \left(\frac{d\rho}{d\gamma}\right)^2 \mathrm{var}(\hat{\rho}).
\end{equation}
It can be shown from (\ref{rho}) and standard relations for the Bessel functions $I_0(\gamma)$ and $I_1(\gamma)$ that
\begin{equation}\label{drhodgamma}
\frac{d\rho}{d\gamma}
= \sqrt{\frac{\pi}{2\gamma}}\,\frac{\exp(-\gamma)}{2I_0(\gamma)}\{1 - A(\gamma)\}.
\end{equation}
Hence,
\begin{equation}\label{MainVar}
\mathrm{var}(\hat{\gamma}_{\text{MOM}})
= \frac{\pi}{8\gamma}\,\frac{\exp(-2\gamma)}{I_0(\gamma)^2}\{1 - A(\gamma)\}^2\,\mathrm{var}(\bar{R}).
\end{equation}
We now quote the following asymptotic variance expressions for $\bar{R}$ in the von Mises case so that Approx~1 can be used. For large $\kappa$, from (4.8.31) of \citet{mardiajupp2000},
\begin{equation}\label{ConcVarR}
\mathrm{var}_1(\bar{R}) = \frac{1}{2n\kappa^2},
\end{equation}
whereas for large $n$, from (4.8.18),
\begin{equation}\label{AsymVarR}
\mathrm{var}_2(\bar{R}) = \frac{1}{n}\left\{1 - A(\kappa)^2 - \frac{A(\kappa)}{\kappa}\right\}.
\end{equation}

Thus we obtain two asymptotic expressions for $\mathrm{var}(\hat{\gamma})$ by substituting either $\mathrm{var}_1(\bar{R})$ from (\ref{ConcVarR}) or $\mathrm{var}_2(\bar{R})$ from (\ref{AsymVarR}) into (\ref{MainVar}), depending on the values of $\kappa$ and $n$.

For CSM,
\[
\hat{\rho}^2_{\text{MOM}} \sim N\!\left(\rho^2,\ \frac{2}{n\,\mathrm{var}(\bar{R})}\right),
\]
and again we may substitute either $\mathrm{var}_1(\bar{R})$ or $\mathrm{var}_2(\bar{R})$.

We now report some experiments showing that the mle is unique and examine the performance of the mle approximations.

\subsection{Numerical experiments}
\label{sec:MLE}
\begin{figure}[!htb]
\centering
\includegraphics[width=\linewidth]{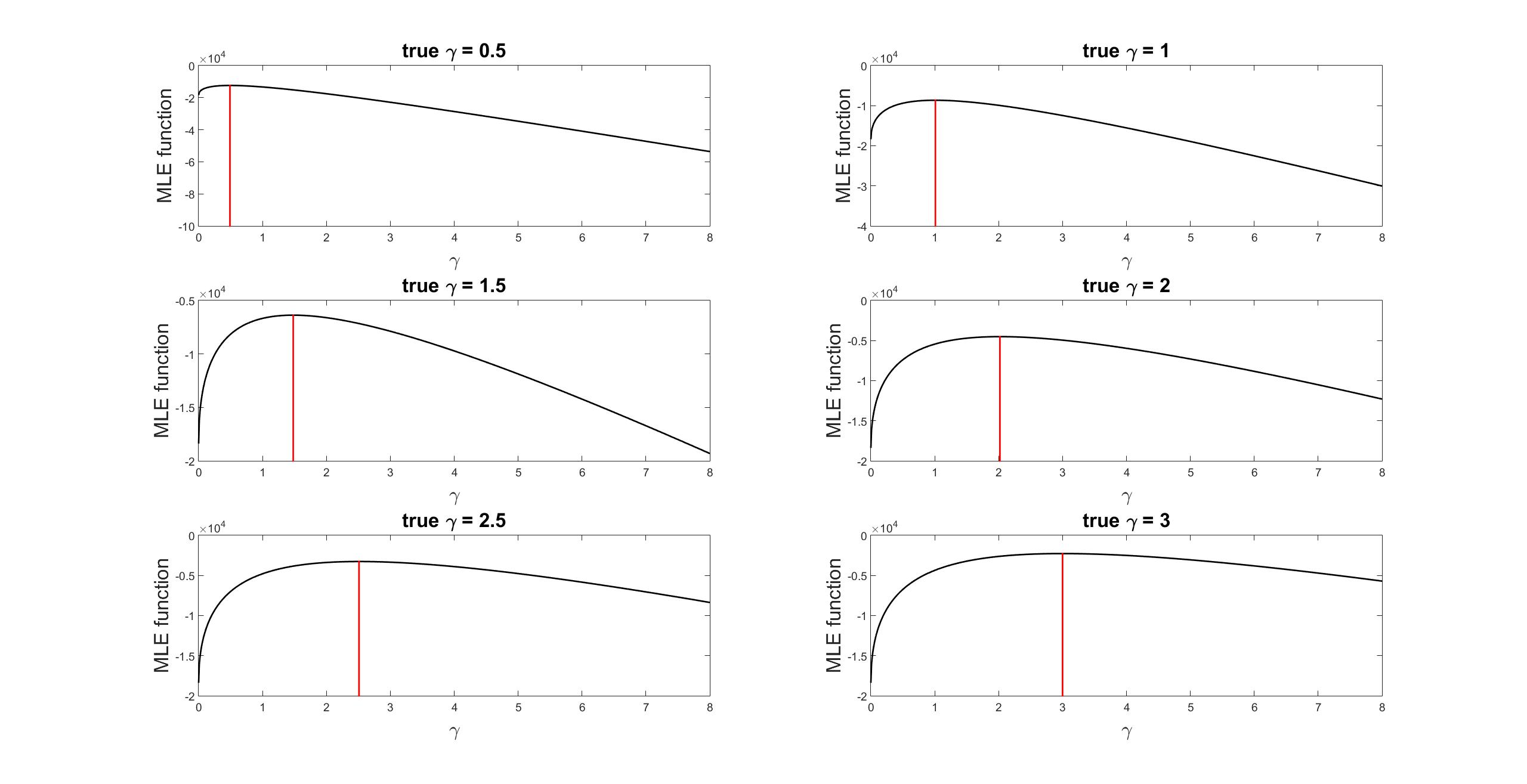}
\caption{Log‑likelihood  (\ref{LLM})  for samples of size \(n=10\) drawn from the PIN distribution with \(\gamma=0.5,1,1.5,2, 2.5, 3\).  Red vertical lines indicate the modes of the respective curves.}
\label{fig:LikelPIN}
\end{figure}

\paragraph{Uniqueness of the MLE of $\gamma$.} To investigate the uniqueness and accuracy of the maximum likelihood estimator \(\hat{\gamma}\), we performed a simulation study.  For each true value \(\gamma\in\{0.5,1,1.5,2,3\}\) we generated samples of size \(n=10\) from the PIN model and computed the log‑likelihood \(\log L(\gamma)\) (see \ref{LLM}  ).  Figure \ref{fig:LikelPIN} displays the log‑likelihood curves for these settings.  Each curve is unimodal and the mode lies very close to the true value of \(\gamma\), supporting the practical uniqueness of the MLE in these cases.

\begin{figure}[!htb]
\centering
\includegraphics[width=.7\linewidth]{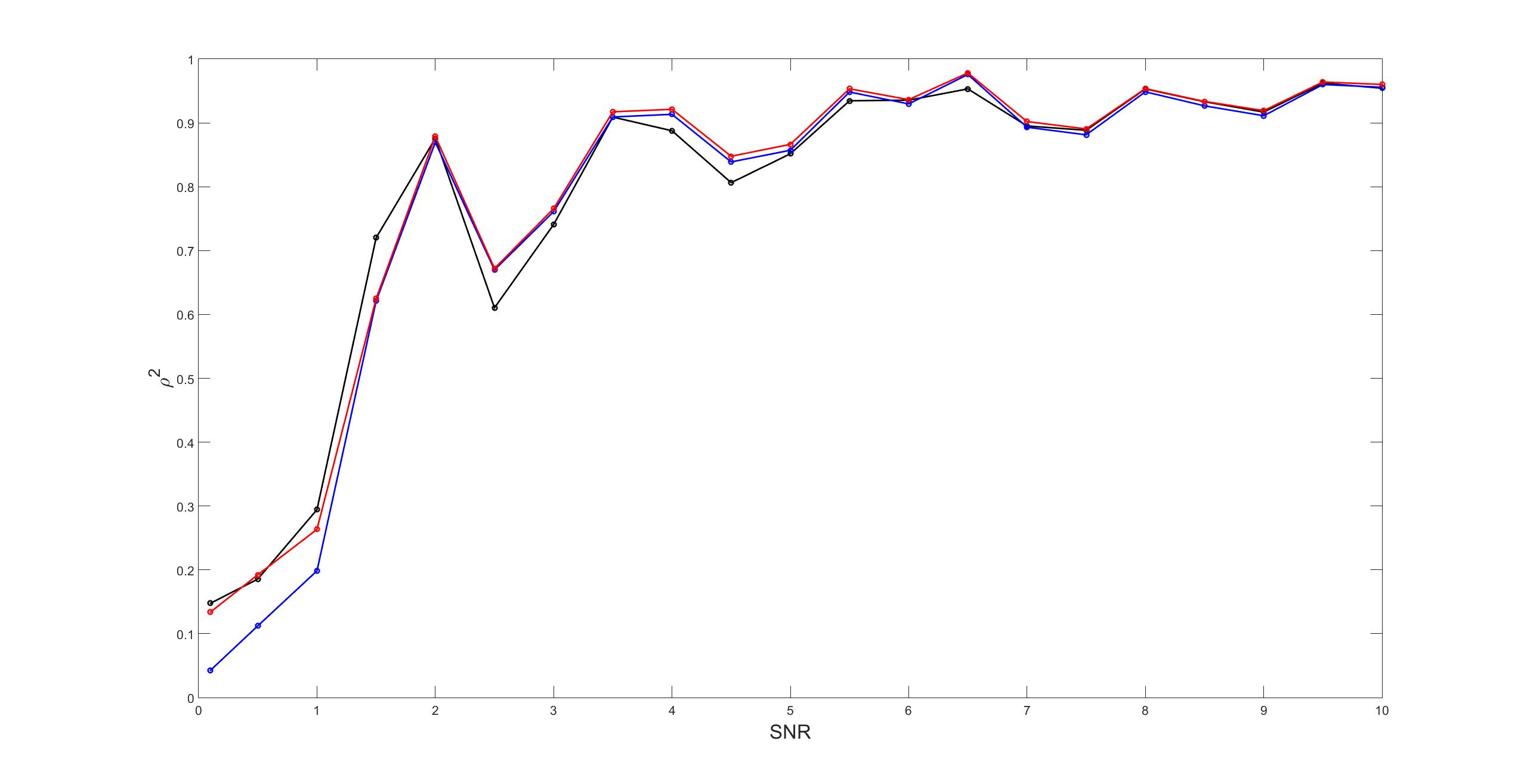}\\[6pt]
\includegraphics[width=.7\linewidth]{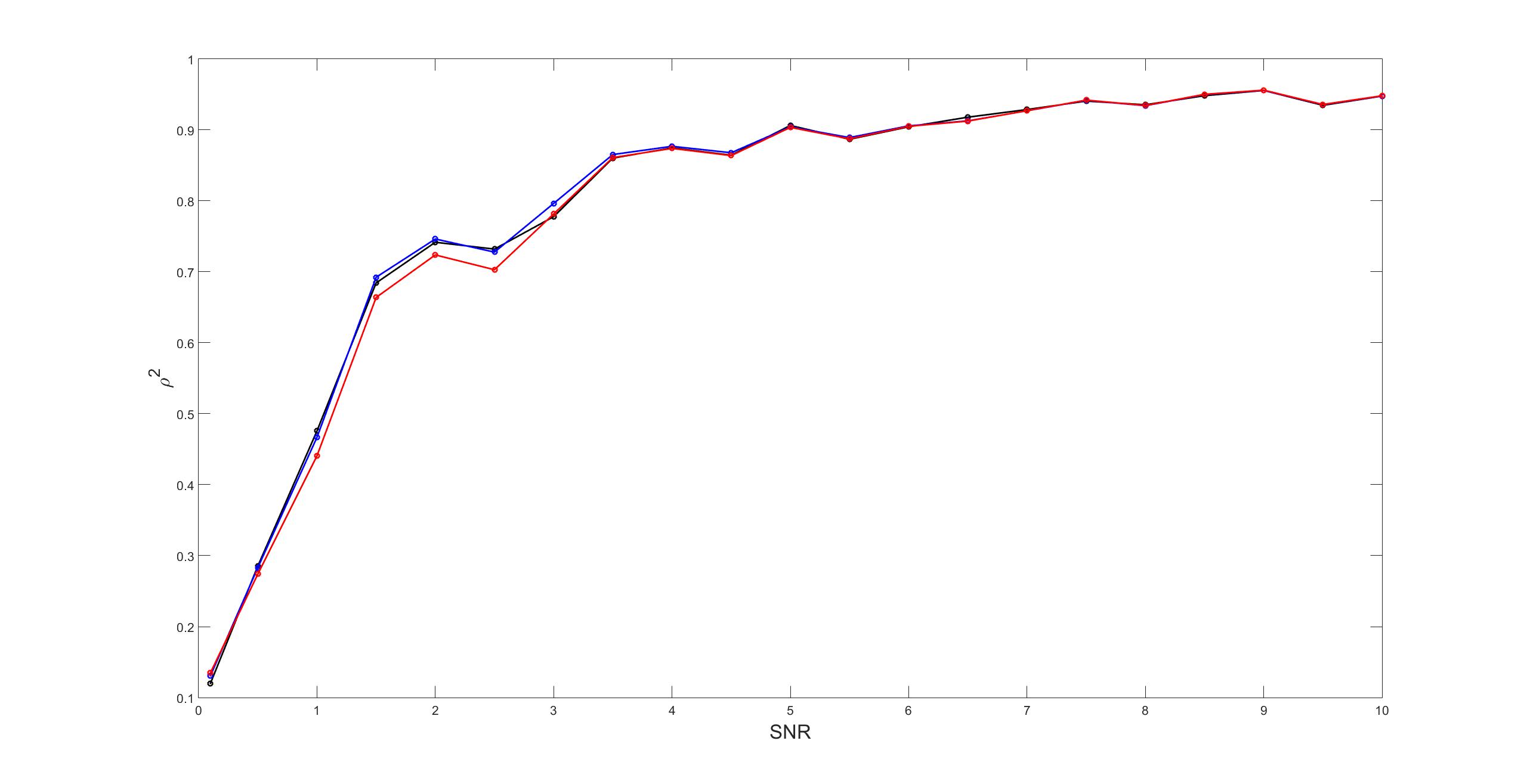}\\[6pt]
\includegraphics[width=.7\linewidth]{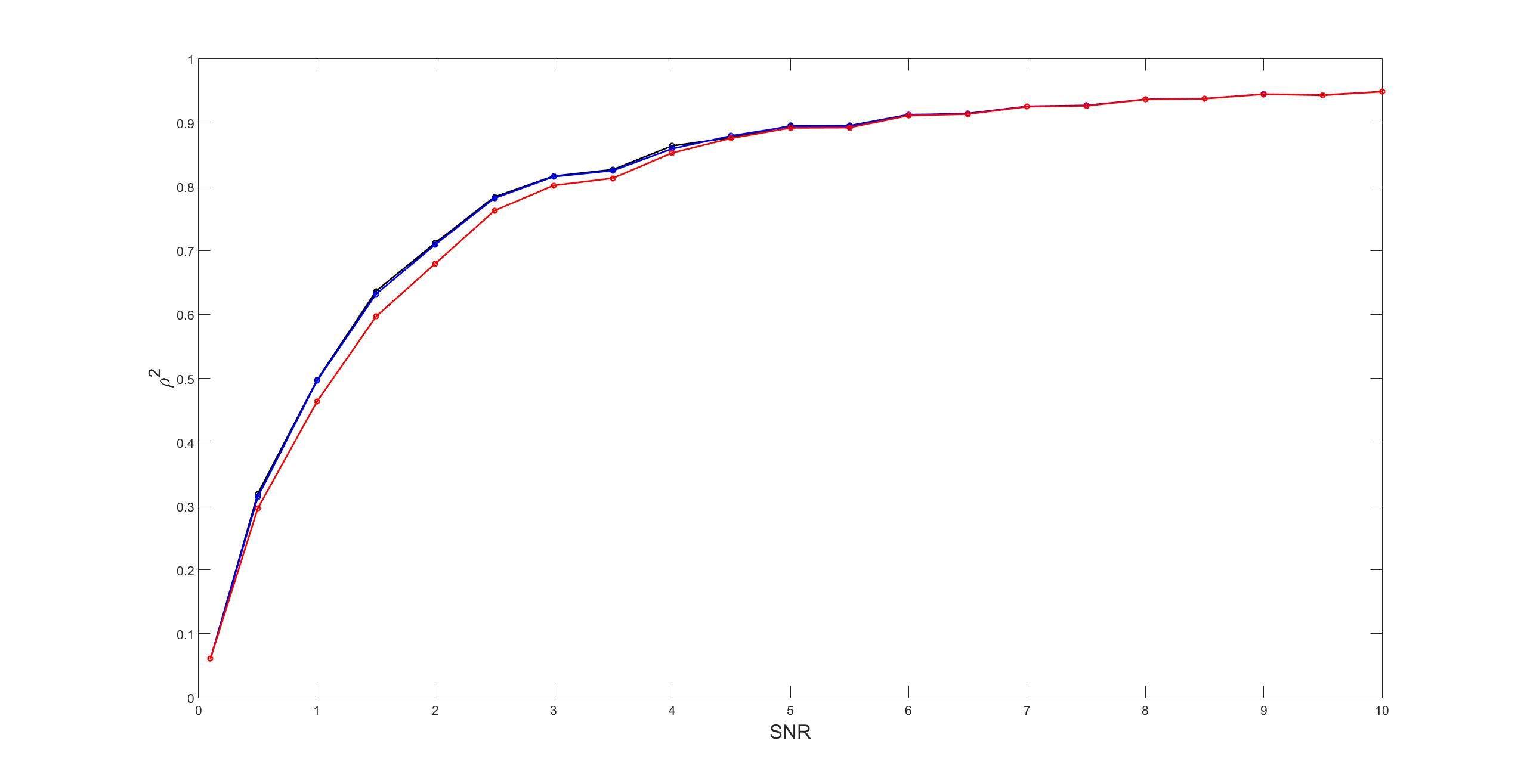}
\caption{Estimates of \(\rho^2\) against SNR (=2$\gamma$) using Approx 1 (blue), Approx 2 (red), and the MLE (black) for sample sizes \(n=10\) (top), \(n=100\) (middle), and \(n=1000\) (bottom).}
\label{fig:MLEn}
\end{figure}
\paragraph{MLE with approximations.}  We compared   Approx 1 and Approx 2 to h the MLE  of  across \(\rho^2\) (CSM)  for a range of SNR (=2$\gamma$) values and sample sizes  \(n=10,100,1000\) .  Figure \ref{fig:MLEn} shows estimates of \(\rho^2\) (CSM) obtained from Approx 1 (blue), Approx 2 (red), and the MLE (black).  Both approximations perform well overall, but their relative accuracy depends on \(n\) and the SNR which we now summarize. 
\begin{itemize}
  \item For large \(n\) (e.g., \(n=1000\)), Approx 1 closely matches the MLE across the SNR range.
  \item For small \(n\) (e.g., \(n=10\)) and low-to-moderate SNR, Approx 2 is slightly closer to the MLE.
  \item In practice the two approximations are similar for most SNR values; differences are most noticeable in the mid‑range SNR and for small sample sizes.
\end{itemize}

 For routine applications with moderate to large sample sizes, Approx 1 is convenient and more accurate.  When sample sizes are small and SNR is in the mid range, Approx 2  provides a modest improvement.  When the highest possible accuracy is required (for example, when making inference in the tails of the distribution), the MLE remains the preferred choice.

\section{Hypothesis Testing and Confidence Intervals}\label{Tests}

\textbf{Testing of Hypotheses.}
Let $\theta$ be distributed as PIN($\mu,\gamma$) and let $\theta_i, i=1, \ldots, n$ be a random sample drawn from this distribution. We want to test the null hypothesis $H_0$ of uniformity against a simple alternative:
\[
H_0:\ \gamma = 0 \qquad \text{vs} \qquad H_1:\ \gamma = \gamma_1.
\]
We can write the log-likelihood ratio (LRT) from (\ref{PIN}) as
\begin{equation}\label{LL1}
- n \log(2\pi) + \sum \log f(\theta_i;\mu,\gamma_1).
\end{equation}

For a composite alternative with $\mu$ unknown,
\[
H_0:\ \gamma = 0 \qquad \text{vs} \qquad H_1:\ \gamma \ne 0,
\]
the log-likelihood ratio is
\begin{equation}\label{LLR}
- n \log(2\pi) + \sum \log f(\theta_i;\hat{\mu},\hat{\gamma}),
\end{equation}
where $\hat{\mu}$ and $\hat{\gamma}$ are the maximum likelihood estimates described in Section~\ref{Sec: Est}. We cannot simplify the LRT statistics in (\ref{LL1}) and (\ref{LLR}); for data, they must be computed numerically.

One way to partially overcome this difficulty is to use a von Mises approximation. For the von Mises case, the LRT test in (\ref{LLR}) reduces to the Rayleigh test, which is uniformly most powerful (see, for example, \citet{mardiajupp2000}). The test is
\begin{equation}\label{Raleigh}
\text{reject } H_0 \quad \text{for large values of } \bar{R}.
\end{equation}
Tables for the von Mises case are available (see \citet{stephens1969tests}). For large $n$, we may use the approximation in (\ref{Unif}); the statistic has a chi-square distribution with 2 degrees of freedom (i.e., an exponential distribution), and the critical value is
\[
\bar{R}^2_{\mathrm{crit}} = \frac{\ln(1/\alpha)}{n}.
\]

For use in Section~\ref{Appl}, note that for $\alpha = 0.05$ and $n = 12$, we have $\bar{R}^2_{\mathrm{crit}} = 0.25$ (as used in Figure~\ref{O1ApplCSM}). From \citet{stephens1969tests}, the exact $5\%$ value of $\bar{R}$ is $0.494$, so $\bar{R}^2 = 0.244$, showing that the approximation is good even for $n=12$.

Also, for large $n$,
\[
E(\bar{R}^2) = \frac{1}{n}, \qquad \mathrm{var}(\bar{R}^2) = \frac{1}{n^2},
\]
leading to the simpler approximation
\begin{equation}\label{UniformNormal}
(n\bar{R}^2 - 1) \sim N(0,1).
\end{equation}
For $5\%$ significance and $n=12$, this approximation gives $\bar{R}^2 = 0.247$, which is slightly better in this case.

Indeed, we can carry out inference for the PIN distribution approximately (especially for large $n$) using Approx~1 or Approx~2 by using the corresponding von Mises solution. This includes confidence intervals for $\gamma$ via $\kappa$, and other inferential procedures, which we now describe. More examples appear in the application section.

\textbf{Confidence Intervals.}
Let us consider confidence intervals for $\gamma$ by using the confidence intervals for $\kappa$ given in \citet{mardiajupp2000}, pp.~150–151, and then applying one of our approximations as described below.

\begin{enumerate}
\item For $\kappa \ge 2$ (equivalently, $\gamma \ge 0.5$), an approximate $(1-\alpha)$ confidence interval for $\kappa$ is
\begin{equation}\label{CIkappa}
(\kappa_l, \kappa_u), \qquad
\kappa_l = \frac{1 + \sqrt{1 + 3a}}{4a}, \qquad
\kappa_u = \frac{1 + \sqrt{1 + 3b}}{4b},
\end{equation}
with
\begin{equation}\label{ab}
a = \frac{n - R}{\chi^2_{n-1;\,1-\alpha/2}}, \qquad
b = \frac{n - R}{\chi^2_{n-1;\,\alpha/2}},
\end{equation}
where $\chi^2_{n-1;\,\alpha/2}$ denotes the upper $\alpha/2$ point of the $\chi^2_{n-1}$ distribution.

For example, using $\kappa = 4\gamma$ for large $\gamma$ (from Approx~1 or Approx~2), we obtain an approximate $(1-\alpha)$ confidence interval for $\gamma$ from (\ref{CIkappa}):
\begin{equation}\label{CISNR}
\frac{1}{4}\kappa_l \;\le\; \gamma \;\le\; \frac{1}{4}\kappa_u,
\end{equation}
where $\kappa_l$ and $\kappa_u$ are given by (\ref{CIkappa}) in terms of $R$.

\item For $\kappa < 2$ (equivalently, $\gamma < 0.5$), we may use the bootstrap (see \citet{fisher1993}, pp.~88–92). Before constructing a confidence interval, however, we should first test the null hypothesis of uniformity, $\kappa = 0$. If the null is accepted, a confidence interval for $\kappa$ is not relevant.
\end{enumerate}

Note that for $\kappa = 2$, we have $\rho = 0.698$ and $\rho^2 = 0.487$, i.e., $\rho^2 \approx 0.5$ to one decimal place. Thus, if we observe $\bar{R}^2 \ge 0.5$, we may use the confidence interval for $\gamma$ given in (\ref{CISNR}).

\section{Applications to Biomedical Signals: EEG Stimulation Study}\label{Appl}

\begin{figure}[!htb]
    \centering
    \includegraphics[width=1\linewidth]{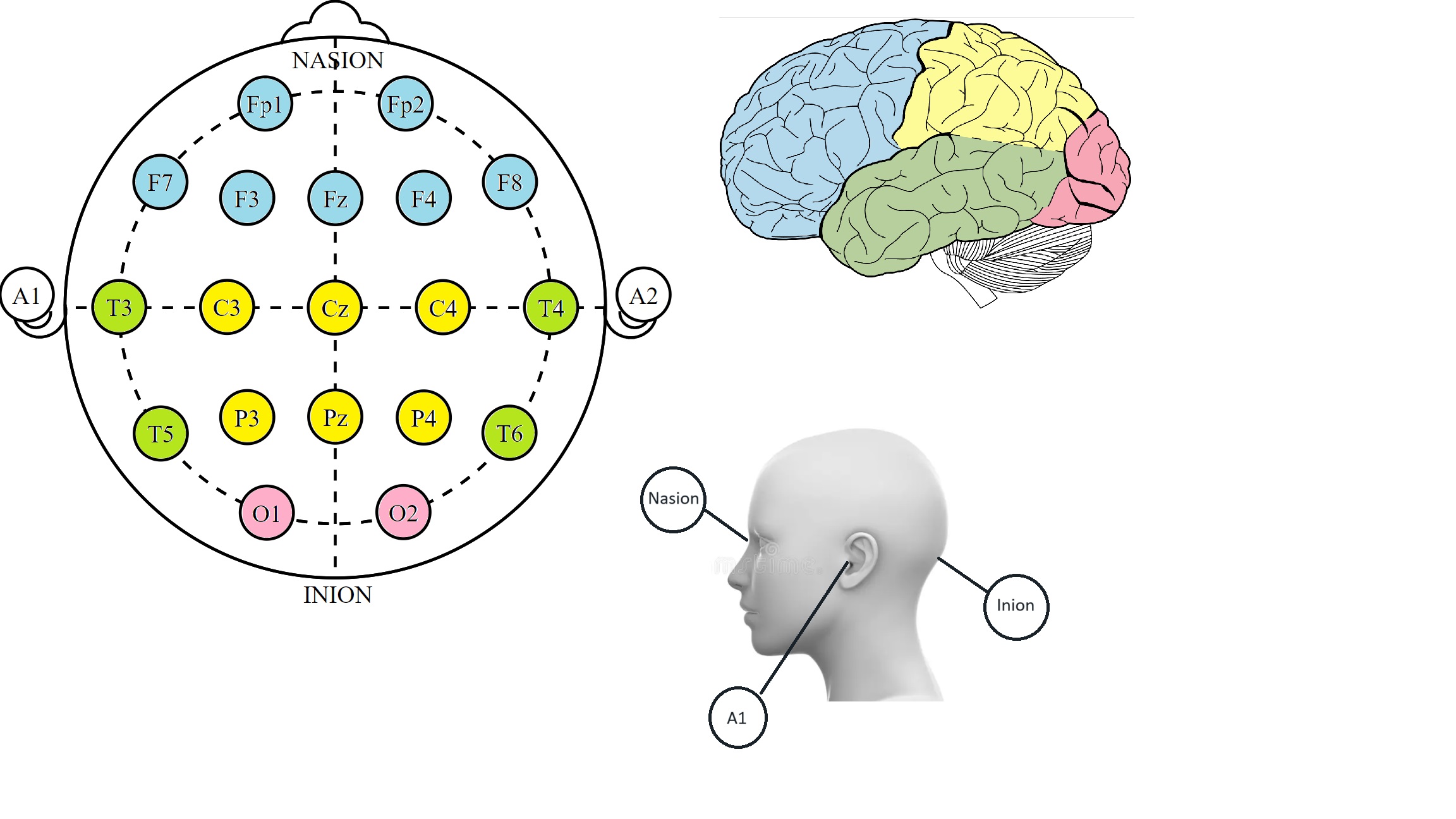}
    \caption{EEG electrode positions in the 10–20 international system. The electrode sites are shown in the same colours as the corresponding underlying lobes of the brain: F (frontal), C (central), P (parietal), O (occipital), and T (temporal). The head diagram indicates the locations of the fiducial points (nasion, left pre‑auricular point, and inion). }
    \label{elect}
\end{figure}

We present an illustrative example of biomedical signals consisting of electroencephalogram (EEG) recordings during a visual type of stimulation called intermittent photic stimulation (IPS). We first describe how the EEG trace data are acquired. In general, electrodes are placed on the scalp according to standardized positions, as shown in Figure~\ref{elect}, which displays the common 10–20 international EEG system. According to convention, the letters indicate the anatomical regions beneath the electrode: O (occipital), T (temporal), C (central), P (parietal), and F (frontal). Odd numbers denote electrodes in the left hemisphere and even numbers those in the right hemisphere.

\begin{figure}[!htb]
    \centering
    \includegraphics[width=1\linewidth]{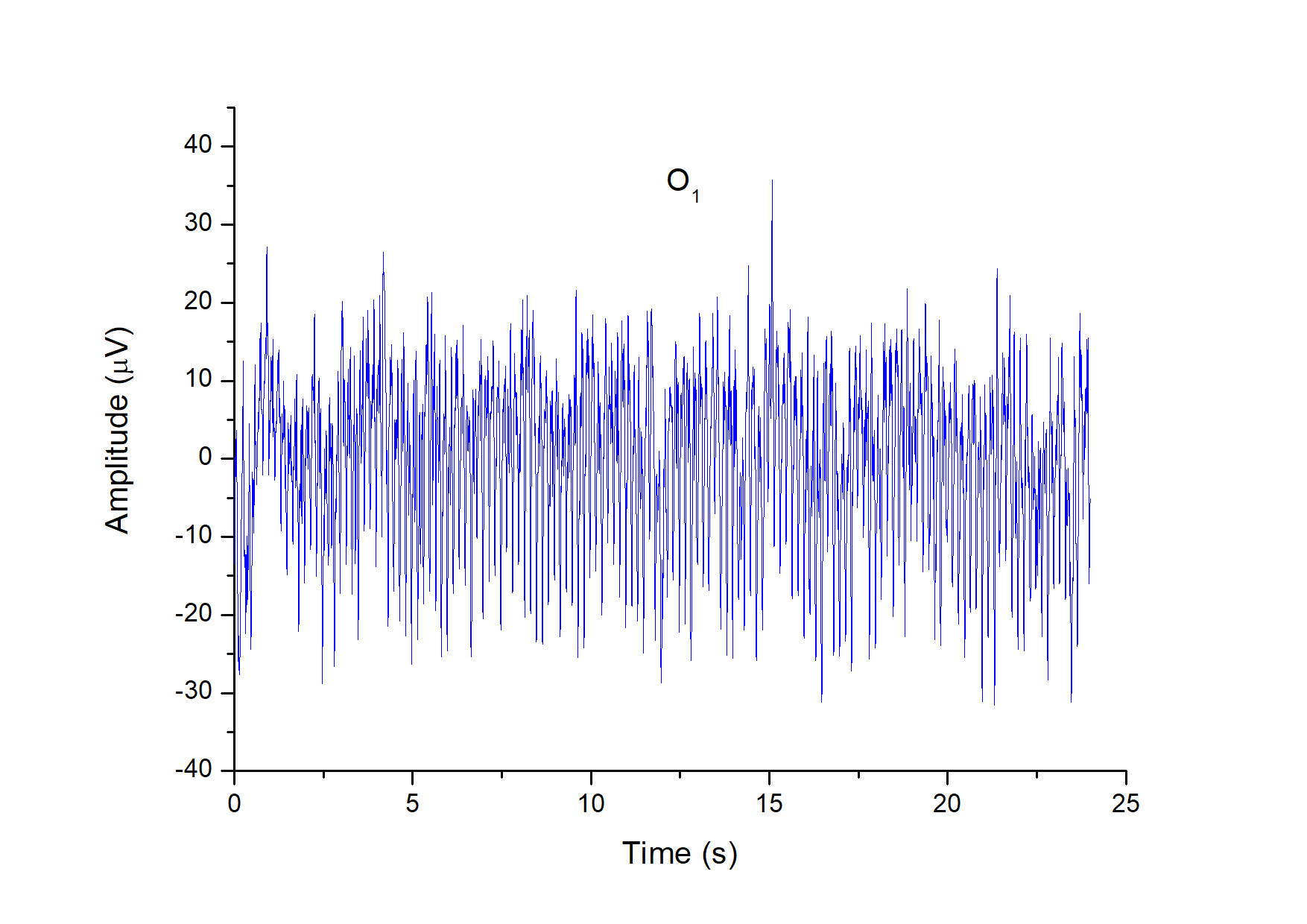}
    \caption{Time series of EEG recorded at electrode $O_1$ for a subject stimulated at 6~Hz with stroboscopic flickering light. x‑axis: time in seconds; y‑axis: scaled voltage.}
    \label{O1TS}
\end{figure}

The data used here were collected in \citet{Mir2002}. These consist of EEG recordings during IPS with stroboscopic light flashing at 6~Hz (six flashes per second) from a subject at two electrodes close to the visual cortex ($O_1$ with O = occipital, and $P_3$ with P = parietal; see Figure~\ref{elect}). The EEG was digitised using a 16‑bit analogue‑to‑digital converter with a sampling frequency of 256~Hz. Each 1‑second signal produced a time series of 256 points. The EEG time series for electrode $O_1$ over 24 seconds is shown in Figure~\ref{O1TS}.

\begin{figure}[!htb]
    \centering
    \includegraphics[width=1\linewidth]{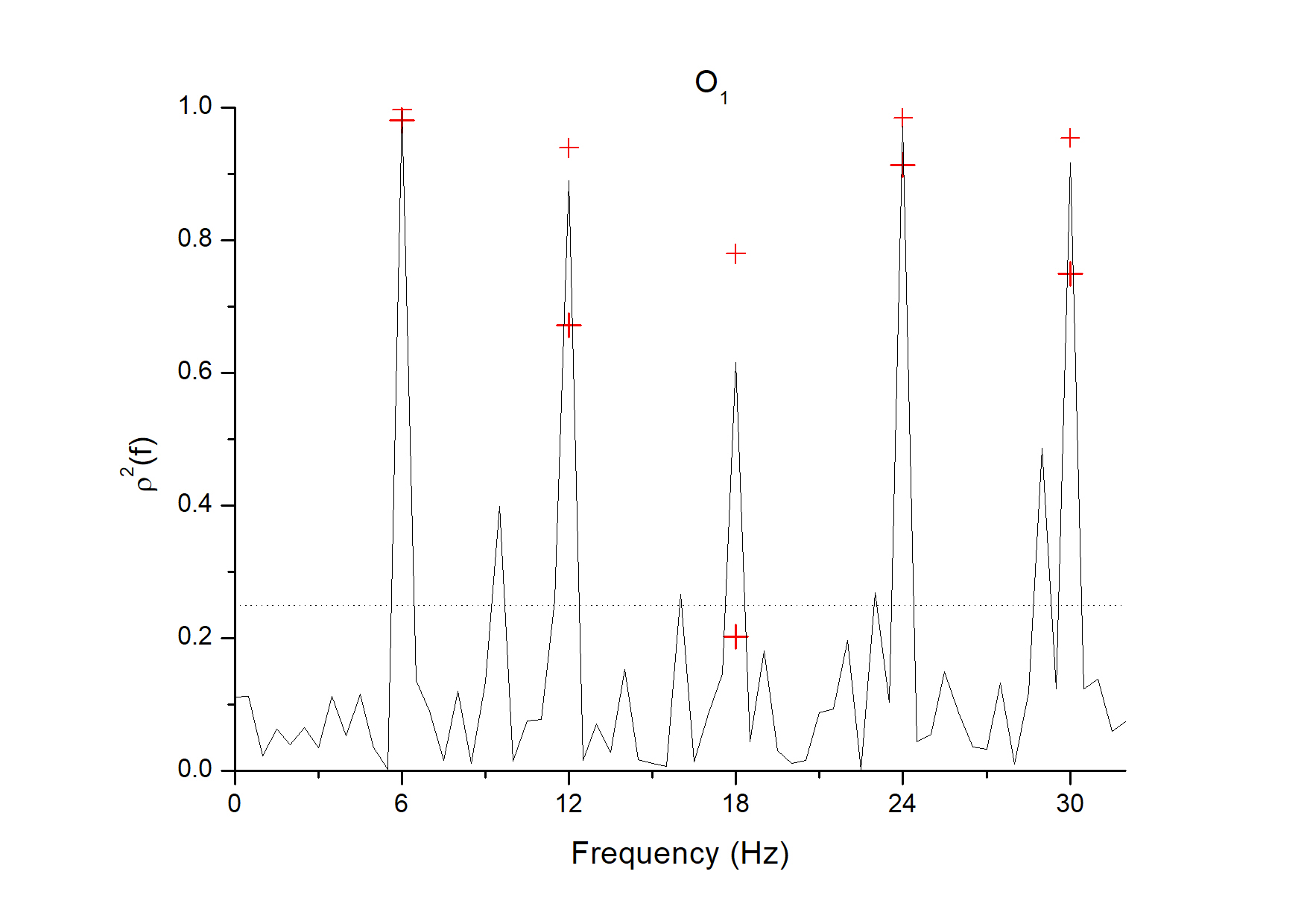}
    \caption{CSM for a subject stimulated at 6~Hz with stroboscopic flickering light at electrode $O_1$. The `+' symbols indicate 95\% confidence limits at the harmonics of stimulation. The horizontal dotted line indicates the critical value under uniformity ($\bar{R}^2 = 0.25$, $\bar{R} = 0.5$).}
    \label{O1ApplCSM}
\end{figure}

We split the time series into consecutive 2‑second segments, yielding $n=12$ sub‑time series, each with 512 points. Figure~\ref{O1TS} displays the entire 24‑second EEG record used for estimating the CSM.
The discrete Fourier transform of each sub‑series was computed using the fast Fourier transform (FFT) algorithm (each sub‑series has $2^{9}=512$ points). The phase of the Fourier transform was then extracted as required in our PIN model. Thus, at each frequency, we obtained 12 phase angles and computed the CSM from (\ref{CSF}). Figure~\ref{O1ApplCSM} shows the CSM for this dataset plotted against frequency (up to 36~Hz). The horizontal dotted line marks the 95\% critical value under uniformity, which from Section~\ref{Tests} is $0.25$. The figure also shows the 95\% confidence intervals at frequencies $6, 12, 18, 24, 30$~Hz obtained from Section~\ref{Tests}.

\begin{figure}[!htb]
    \centering
    \includegraphics[width=1\linewidth]{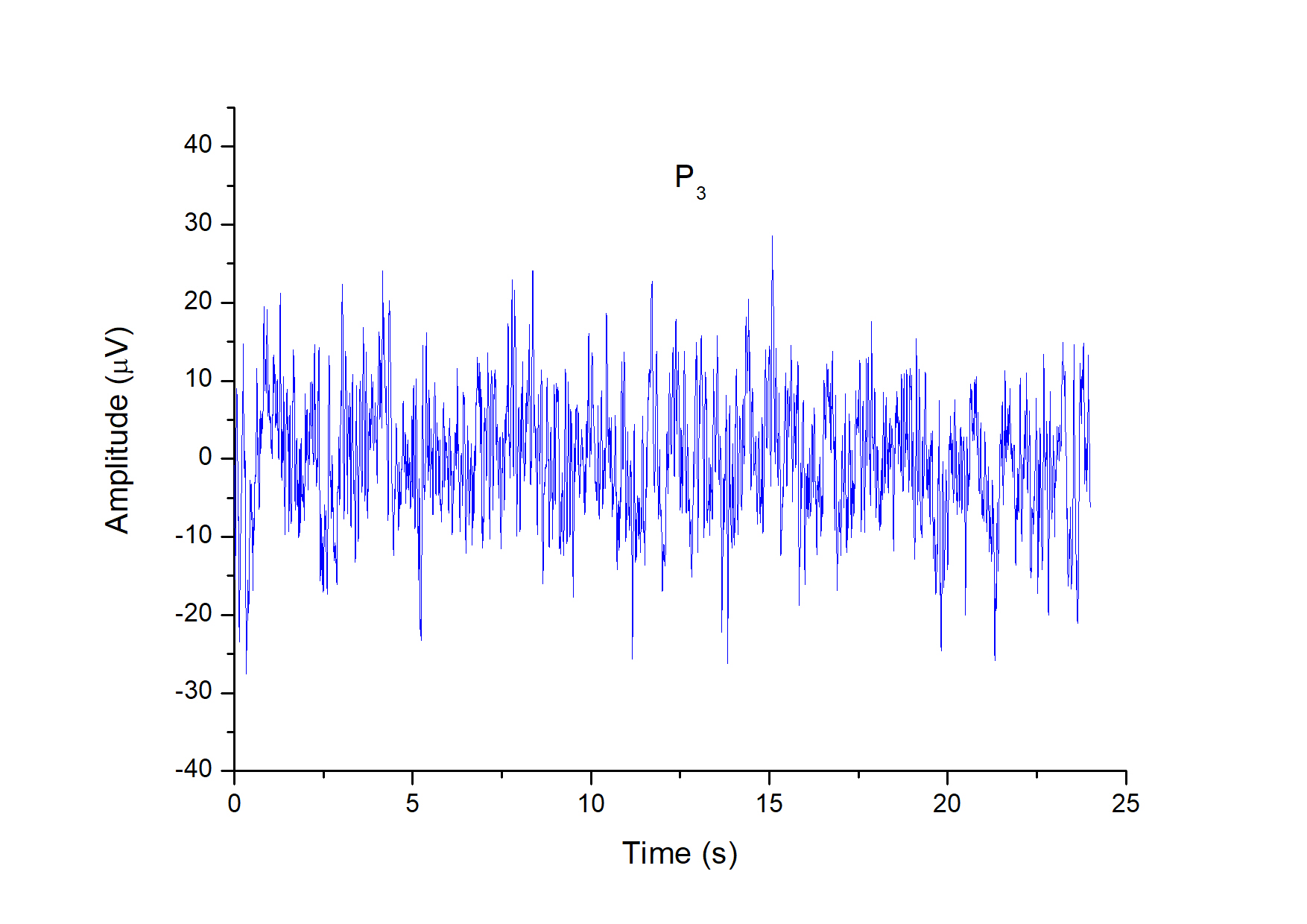}
    \caption{Time series of EEG recorded at electrode $P_3$ for a subject stimulated at 6 Hz with stroboscopic flickering light, x‑axis: time in seconds; y‑axis: scaled voltage.}
    \label{P3TS}
\end{figure}

\begin{figure}[!htb]
    \centering
    \includegraphics[width=1\linewidth]{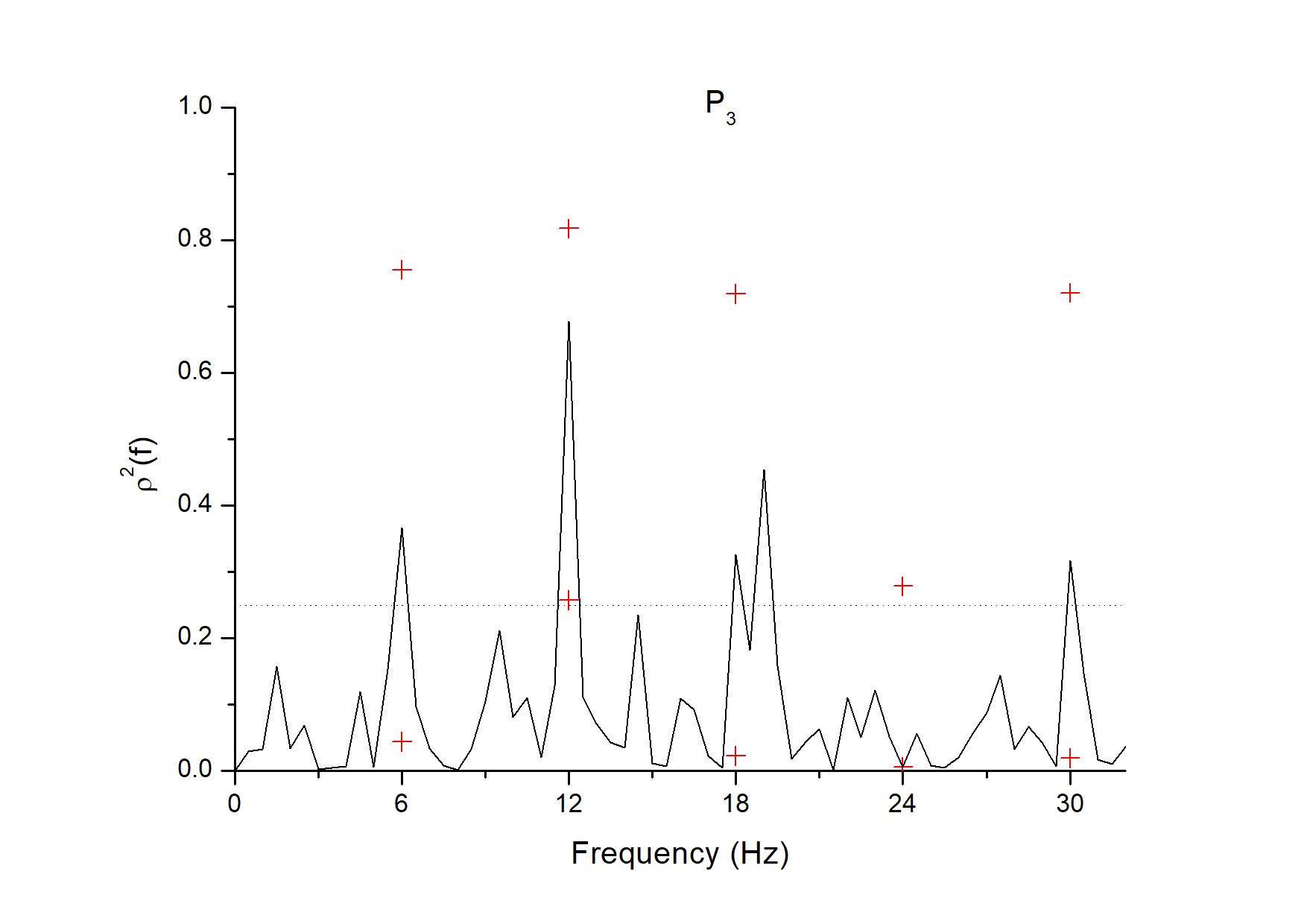}
    \caption{CSM for a subject stimulated at 6 Hz with stroboscopic flickering light at electrode $P_3$. The `+' symbols indicate 95\% confidence limits at the harmonics of stimulation. The horizontal dotted line indicates the critical values under uniformity ($\bar{R}^2 = 0.25$, $\bar{R} = 0.5$).}
    \label{P3ApplCSM}
\end{figure}

For comparison, we also present results for electrode $P_3$. Figure~\ref{P3TS} shows the corresponding time series, and Figure~\ref{P3ApplCSM} shows the CSM against frequency.  Visually, the two CSM plots (Figures~\ref{O1ApplCSM} and \ref{P3ApplCSM}) differ markedly at the peaks, except at 12~Hz. As an illustration, we test the hypothesis of equality of the two concentration (SNR) parameters at frequency $f = 6$~Hz. The data with $n=12$ are given in Table~\ref{tab:phase-data}, and summary statistics appear in Table~\ref{tab:phase-summary}.

\begin{table}[ht]
\centering
\caption{Phase angles of $O_1$ and $P_3$ at 6 Hz, $n=12$}
\label{tab:phase-data}
\setlength{\tabcolsep}{3pt}
\renewcommand{\arraystretch}{1.25}
\setlength{\extrarowheight}{1pt}
\scriptsize
\begin{tabular}{l r r r r r r r r r r r r}
\hline
O1 & -2.2032 & -1.9798 & -2.0625 & -2.2151 & -2.2389 & -2.0569 & -2.2505 & -2.1924 & -2.1404 & -2.1541 & -2.1244 & -2.1647 \\
P3 &  2.1879 & -0.2305 & -1.6763 & -1.7409 & -2.8771 & -1.9322 &  2.9193 &  2.8651 & -3.0499 & -1.9783 &  3.0112 & -2.7492 \\
\hline
\end{tabular}
\normalsize
\end{table}

\begin{table}[ht]
\centering
\caption{Summary statistics for the phase angle data of $O_1$ and $P_3$, $n=12$}
\label{tab:phase-summary}
\small
\begin{tabular}{l c c c}
\hline
\textbf{Statistic} & \textbf{$O_1$} & \textbf{$P_3$} & \textbf{$O_1+P_3$} \\
\hline
$\bar{C}$      & $-0.5445$ & $-0.5370$ & $-0.5408$ \\
$\bar{S}$      & $-0.8351$ & $-0.2799$ & $-0.5575$ \\
$\bar{\theta}$ & $237^\circ$ & $208^\circ$ & $226^\circ$ \\
$\bar{R}$      & $0.997$   & $0.606$   & $0.777$ \\
CSM            & $0.9939$  & $0.3667$  & $0.6037$ \\
\hline
\end{tabular}
\end{table}

Let us assume that the samples for $O_1$ and $P_3$ come from PIN($\mu_1,\gamma_1$) and PIN($\mu_2,\gamma_2$), respectively.
First, we report using Section  \ref{Sec: Est} that the hybrid estimates of $\gamma_1$ and $\gamma_2$ are 41.24 and 0.29, respectively, confirming that $O_1$ is highly concentrated in comparison with $P_3$.

As an illustration, we now test the hypothesis
\[
H_0:\ \gamma_1 = \gamma_2 \qquad \text{vs} \qquad H_1:\ \gamma_1 \ne \gamma_2.
\]
Let $\lambda$ be the  likelihood ratio  (Section \ref{Tests}) . We find $-2\log\lambda = 43.7$, which is approximately $\chi^2_2$ and therefore highly significant. Assuming a von Mises model, we have
\[
F_{11,11} = \frac{1 - \bar{R}_2}{1 - \bar{R}_1} = 101,
\]
again highly significant.
Further, the 95\% confidence interval for the CSM  from (\ref{CISNR}) of  Section \ref{Tests} of $O_1$ is $(0.9810,\ 0.9967)$, and the 95\% confidence interval for the CSM of $P_3$ is $(0.0507,\ 0.7652)$. This again confirms that $O_1$ is highly concentrated, whereas $P_3$ is much more diffuse.

%\begin{figure}[!htb]
    %\centering
   %\includegraphics[width=1\linewidth]{O1P3T5_Time_Series.jpg}
   %\caption{Time series of three EEG records collected at electrodes $O_1$, $P_3$, and $T_5$ for a %subject stimulated at 6 Hz with stroboscopic flickering light at the three electrodes.}
  %  \label{TS}
%\end{figure}

%\begin{figure}[!htb]
 %   \centering
%    \includegraphics[width=1\linewidth]{FigEEG.JPG}
 %   \caption{CSM for a subject stimulated at 6 Hz with stroboscopic flickering light at (a) $O_1$, %(b) $P_3$, and (c) $T_5$. '+' indicates confidence limits (95\%) at the harmonics of stimulation %and the horizontal dotted lines indicate the critical values which is $\bar{R}^2=0.25,\bar{R}=0.5 %$}, see Text
 %   \label{ApplCSM}
%\end{figure}

 \section{Discussion}\label{Disc} 

In this paper, we have developed a phase–based framework for analysing EEG signals using the projected isotropic normal (PIN) distribution. Starting from the Fourier representation of EEG traces, we identified the phase as the key quantity for detecting stimulus–related responses and showed that its distribution arises naturally as a projected isotropic normal model. We derived population moments of the PIN distribution and proposed two von Mises approximations, both of which perform well across a wide range of concentration parameters. These results enabled practical inference for the Component Synchrony Measure (CSM), including confidence intervals and hypothesis tests.

We provided estimation procedures for the model parameters, including approximate closed‑form estimators and score‑matching methods, and we developed testing procedures under the PIN framework, including tests for uniformity of phase and related hypotheses.

The application to EEG data under flash stimulation illustrated how the proposed methodology can be used in practice. These tools allow us to assess whether the observed phase distribution departs from uniformity, as would be expected under a stimulus‑related response, and to compare synchrony across conditions using measures such as CSM.

The framework developed here opens several avenues for future work, particularly when moving beyond the single‑electrode setting of our illustrative application. Although we treated the $O_1$ and $P_3$ samples as independent, they are of course correlated because they originate from the same subject. Joint modelling of multiple electrodes is therefore a natural next step. In such settings, the phase angles lie on a torus, and one extension of the projected normal distribution to toroidal data is given by \citet{Kato1925HypertoroidalCopula}. Moreover, EEG recordings exhibit spatial–temporal dependence across electrode sites, and methods for analysing such spatial–temporal data are discussed in \citet{KentMardia2022}; these would be relevant for future developments of our approach.

Although our application focused on EEG data recorded under flash stimulation,\\ phase‑based analyses arise in many other experimental paradigms. For example, \citet{Busch2009} and \citet{BuschVanRullen2010} use the instantaneous phase of spontaneous EEG to study visual attention. This suggests that the PIN framework and the associated inference for CSM may be broadly applicable across cognitive and biomedical neuroscience.

For inference, an alternative to the classical approach adopted here is Bayesian analysis, following \citet{NunezGutierrez2005}. Such methods may be particularly useful when incorporating prior information or when modelling multiple electrodes simultaneously, and they offer a flexible direction for future work.
\section{Acknowledgments}

Kanti Mardia wishes to thank John Kent, Faisal Mushtaq, and  Xiangyu Wu for their helpful comments.

%%%%  

%%%
%%%
\bibliographystyle{rss}
\bibliography{ref_PDWC}

\appendix
\section{Appendix 1:  Deeper Examination of  the EEG Model} \label{Sec: Appendix1Model}
\label{Model}
We give a full  characterization of the EEG model outlined for flash stimulation in Section \ref{Sec: PIN}  using the fundamentals of signal processing and impulse trains. This characterization is implicit in the literature but  has not been reported in any publication.

The intermittent photic stimulation produced by flashes (used in the application in Section \ref{Appl} can be modeled as a periodic impulse train (an evenly spaced series of Dirac delta functions). If the time between consecutive impulses is \(T\), then \(T\) is the stimulation period and the stimulation frequency is \(f_0 = 1/T\).

\begin{figure}[!htb]
\centering
\includegraphics[width=5.5cm]{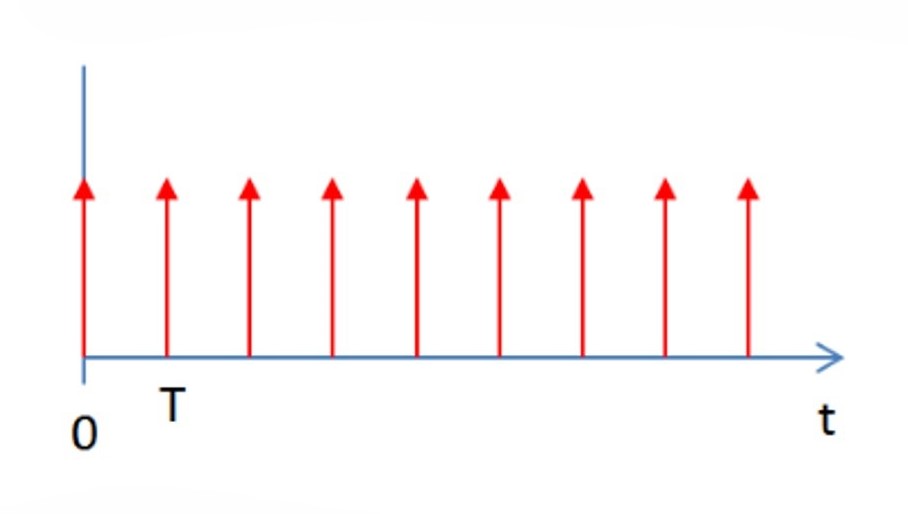}
\hfill
\includegraphics[width=6.0cm]{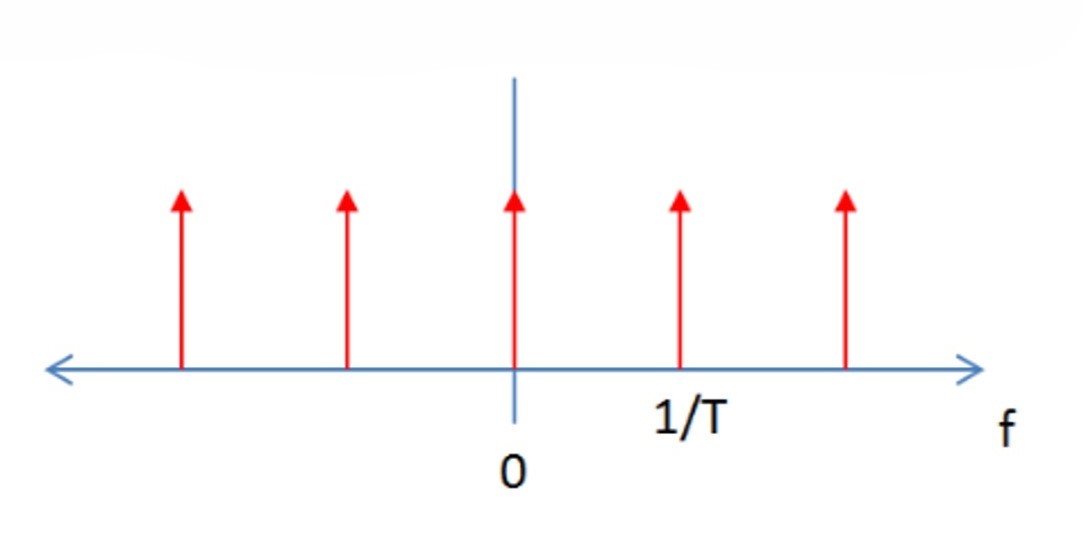}
\caption{Left: impulse train (Dirac deltas) with spacing \(T\). Right: magnitude of its discrete Fourier transform showing spectral impulses with spacing \(1/T\).}
\label{fig:impulse}
\end{figure}

The Fourier transform of a periodic impulse train is another impulse train in the frequency domain with impulses located at integer multiples of the fundamental frequency \(f_0\)  (see below). For example, stimulation at \(6\ \mathrm{Hz}\) corresponds to \(T = 1/6\ \mathrm{s}\) (six flashes per second) and produces spectral impulses spaced by \(6\ \mathrm{Hz}\). That is, the spectrum contains impulses at integer multiples of the fundamental frequency \(f_0=1/T\).
Figure \ref{fig:impulse} shows impulse train (Dirac deltas) with spacing \(T\) (in Left) in time domain and its   corresponding  impulse train with spacing \(1/T\) in frequency domain (in Right). Both are  impulse trains.
\begin{figure}[!htb]
\centering
\includegraphics[width=12cm]{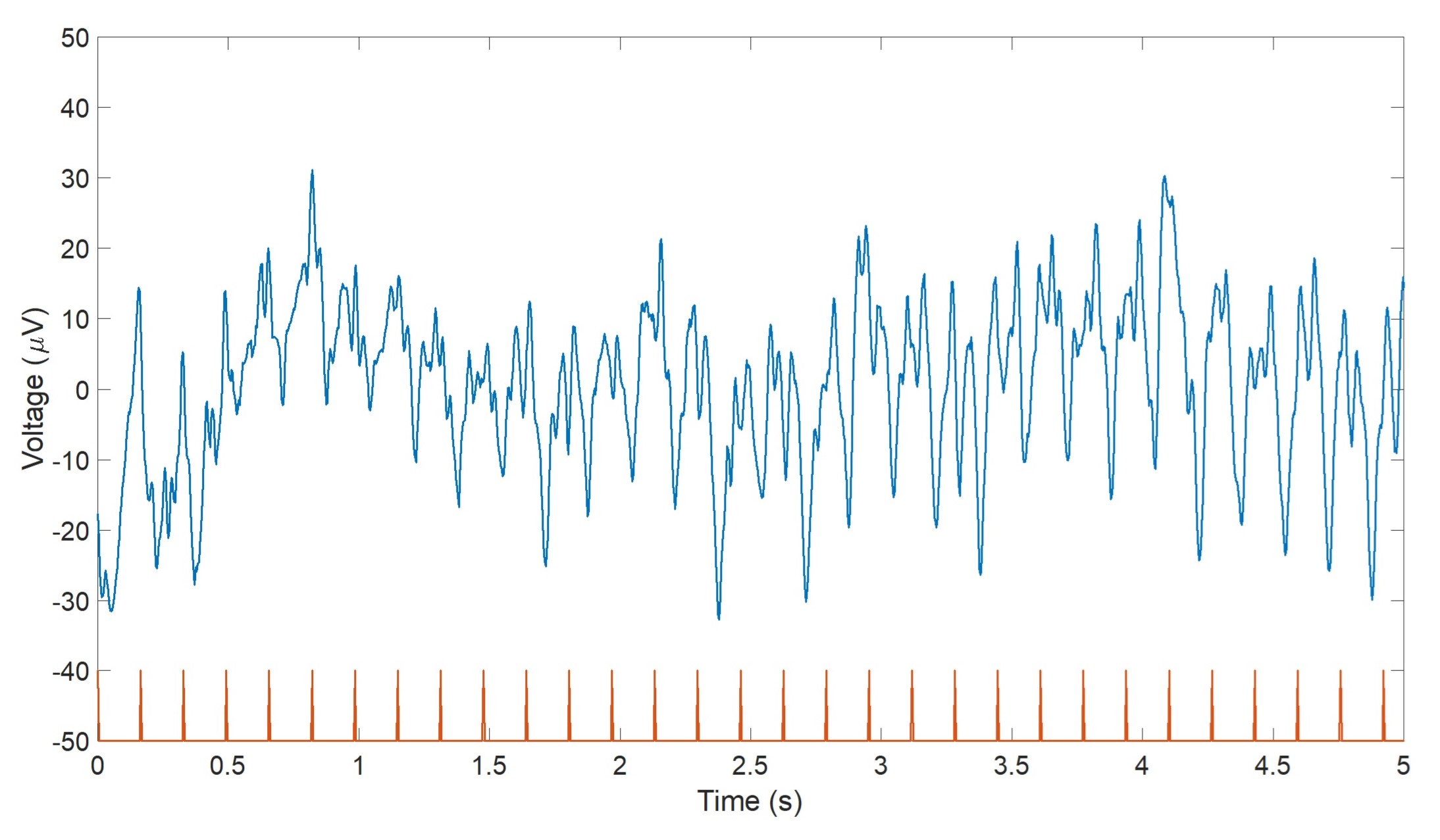}
\caption{Five-second EEG time series from a 13-year-old subject recorded at electrode O1 during 6 Hz photic stimulation. Red marks indicate stimulus times.}
\label{fig:EEG_timeseries}
\end{figure}
\subsection{Impulse trains and the additive linear error model
} 
\begin{figure}[!htb]
\centering
\includegraphics[width=15cm]{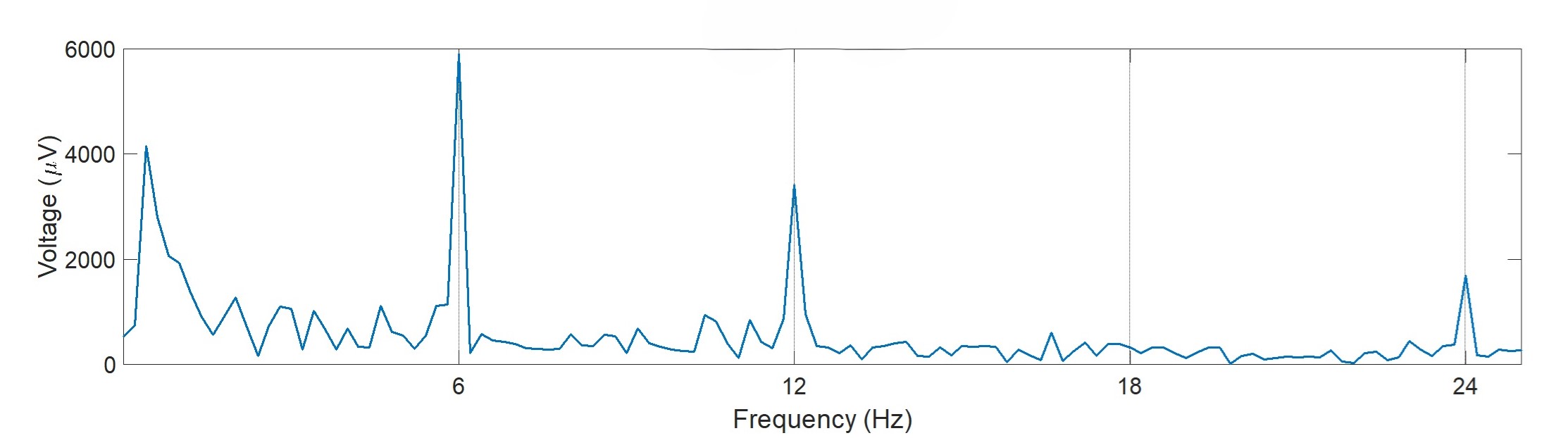}
\caption{Magnitude of the discrete Fourier transform of the 5-second EEG segment. Vertical lines mark the stimulation frequency (6 Hz) and its harmonics.}
\label{fig:EEG_DFT}
\end{figure}

We now give  more details of the output relevant to our EEG application of Section 8 \ref{Appl}. Figure \ref{fig:EEG_timeseries}  shows a five-second EEG time series from a 13-year-old subject recorded at electrode O1 during 6 Hz photic stimulation where  red marks indicate stimulus times st 
time at the step of 1/6 seconds. The corresponding Figure \ref{fig:EEG_DFT} shows the magnitude of the discrete Fourier transform of the 5-second EEG segment where the  vertical lines mark the stimulation frequency (6 Hz) and its harmonics.

{\bf Deterministic Impulse Train }
An ideal continuous-time impulse train with period \(T\) in the {\bf time domain} is
\begin{equation} \label{MathImpulseT}
s(t)=\sum_{n=-\infty}^{\infty} A_n\,\delta(t-nT),    
\end{equation}
where \(\delta(\cdot)\) is the Dirac delta, \(A_n\) is the amplitude of the \(n\)-th impulse (constant or random), and impulses occur at \(t=nT\).
In the {\bf frequency domain},  the Fourier transform of \(s(t)\) is a comb of impulses:
\begin{equation} \label{MathImpulseFreq}
S(f)=\frac{1}{T}\sum_{k=-\infty}^{\infty} A_k\,\delta\!\left(f-\frac{k}{T}\right),  
\end{equation}
where $f$ is the frequency.
\subsection*{Additive  error model: linear input-output system}
We model the observed signal (output) as the impulse train (input) corrupted by additive noise:
\begin{equation} \label{NoiseImpulseT}
x(t)=s(t)+\varepsilon(t),   
\end{equation}
where \(\varepsilon(t)\) is additive Gaussian noise with \(\varepsilon(t)\sim\mathcal{N}(\mu,\sigma^2)\) (applied point-wise in time).
In {\bf frequency-domain},
taking Fourier transforms gives
\begin{equation} \label{NoiseImpulseFreq}
Y(f) = S(f) + N(f)   
\end{equation}
where \(N(f)\) is complex Gaussian noise, \(N(f)\sim\mathcal{CN}(\mu_N,\sigma_N^2)\). The real and imaginary parts of \(N(f)\) are independent normals:
\(\operatorname{Re}N(f),\operatorname{Im}N(f)\sim\mathcal{N}(\mu_N,\sigma_N^2)\). The model (\ref{NoiseImpulseFreq} ) given at  (\ref{spect})  with its discrete Fourier representation and $S(f)$ with deterministic amplitude \(\mu\) at a given frequency.

{\bf Signal-to-Noise Ratio}
Define frequency-specific SNR as
\[
\mathrm{SNR}(f)=\frac{\mathbb{E}\big[|S(f)|^2\big]}{\mathbb{E}\big[|N(f)|^2\big]}.
\]
For complex Gaussian noise with variance \(\sigma_N^2\) per real component,
\(\mathbb{E}[|N(f)|^2]=2\sigma_N^2\), so
\[
\mathrm{SNR}(f)=\frac{\mathbb{E}\big[|S(f)|^2\big]}{2\sigma_N^2}.
\]
With  \(S(f)\) = \(\mu\) at a given frequency, this reduces to 
\begin{equation} \label{SNR}
\mathrm{SNR}(f)=\mu^2/(2\sigma_N^2),   
\end{equation}
%\(\mathrm{SNR}(f)=\mu^2/(2\sigma_N^2)\).
which is the SNR value  (\ref{Eq: SNR}) .
\subsection{Phase angles of EEG in flash stimulation}
This impulse–train plus noise model captures the principal features of photic stimulation in EEG: a comb-like spectrum at the stimulation frequency and its harmonics, superimposed on broadband noise that can mask or distort the spectral peaks. The model is therefore useful for simulation, for understanding the spectral signatures of stimulation, and for designing detection or filtering strategies based on SNR. Flashes interact with the brain's ongoing rhythms in a time-specific manner: the instantaneous phase of an oscillation at stimulus onset determines neuronal excitability, making phase measurements essential for explaining and predicting the neural and perceptual effects of brief visual stimuli \citep{Busch2009,BuschVanRullen2010}. Furthermore, Notbohm and Herrmann \citep{NotbohmHerrmann2016}, together with Mathewson et~al. \citep{Mathewson2009}, provide compelling evidence that the brain’s responsiveness to visual flash stimulation is governed primarily by the phase of ongoing oscillations—far more than by their amplitude—demonstrating that phase alignment at stimulus onset determines both the strength of neural entrainment and the likelihood of conscious perceptual awareness. Thus, to sum up, the PIN distribution for the phase-angle data in Section \ref{} is appropriate.
\subsection{Origin of CSM}
\label{subsec:CSM-origin}
We now  discuss the model given by (\ref{NoiseImpulseT}) in the discrete time domain and its corresponding Fourier representation 
(\ref{NoiseImpulseFreq}) in the discrete Fourier domain and  recast the method of \citet{Fried1984} into a model framework. 
Below we give a representation for the observed EEG (in time domain and frequency domain)  and  show how the phase statistics used by \citet{Fried1984} lead naturally to the CSM measure. 

Assume a real-valued discrete time series \(x(t)\), \(t=1,\dots,n\), with \(n\) even corresponding to (\ref{NoiseImpulseT}).  A convenient finite Fourier series representation that isolates the sinusoidal components is
\begin{equation}\label{SNRmodel}
x(t)=b_0+\sum_{j=1}^{n/2-1}\bigl[a_j\sin(2\pi j t/n)+b_j\cos(2\pi j t/n)\bigr]+b_{n/2}\cos(\pi t),
\end{equation}
where the coefficients \(\{a_j,b_j\}\) determine the contribution at each Fourier frequency.  We model the coefficients at a given frequency \(j\) as
\begin{equation} \label {ModelSupp} 
  Y_j =  a_j + i b_j = r_j e^{i\theta_j},
\end{equation}
the phase \(\theta_j\) is then a circular random variable whose distribution under the signal-plus-noise model is the PIN model used in our theoretical development.  \citet{Fried1984} focused on the  \(\bar{R}^2\) or CSM (using  sample circular variance as \(1-\bar{R}^2\)  but not the standard variance \(1-\bar{R}\)) computed from the collection of phase angles \(\{\theta_j\}\). The CSM statistic measures phase synchrony (concentration) across trials or channels for a given Fourier coefficient and is therefore useful as a criterion for distinguishing signal from uniform (noise) phase distributions.

In our implementation, we compute the discrete Fourier transform of the discretized EEG \(x(t)\),
\begin{equation}\label{FFT}
 Y_j = r_j e^{i\theta_j}=\sum_{t=1}^n x(t)\,e^{-i2\pi jt/n},\qquad j=1,\dots,n,   
\end{equation}
and then obtain   \(\bar{R}^2\) (CSM)  from the phases \(\{\theta_j\}\) at the frequency of interest.  
\paragraph{Remark.} As a validation of the PIN approximation, the empirical behaviour of the observed resultant for EEG data closely matches the theoretical distribution under the PIN model; see \citet{Mir2003} for related results.

\section{Appendix 2: Moments}\label{Sec: Appendix2Moment}

We now give a proof of the moment formulae (\ref{eq:sinMoment}) and (\ref{eq:cosMoment}).  
Without loss of generality, in our model (\ref{polar}) we take $\sigma = 1$ and $\mu_1 = 2\sqrt{\gamma}$, $\mu_2 = 0$. Then the joint density of $(r,\theta)$ in (\ref{polar}) is
\begin{equation}\label{jointpdf}
    f(r,\theta;\gamma)
    = \frac{\exp(-2\gamma)}{2\pi}\, r \exp(-r^2/2)
      \exp\{2\sqrt{\gamma}\, r\cos\theta\}.
\end{equation}

Hence,
\begin{equation}\label{sine}
E(\sin p\theta)
= \frac{\exp(-2\gamma)}{2\pi}
  \int_0^{\infty} r \exp(-r^2/2)\, a_1(r)\, dr,
\end{equation}
where
\begin{equation}\label{sineInt}
a_1(r)
= \int_0^{2\pi} (\sin p\theta)\,
    \exp\{2\sqrt{\gamma}\, r\cos\theta\}\, d\theta.
\end{equation}

The integrand in $a_1(r)$ is an odd function of $\theta$, so $a_1(r)=0$.  
This proves (\ref{eq:sinMoment}).

To prove (\ref{eq:cosMoment}), note that
\begin{equation}\label{Ncosine}
E(\cos p\theta)
= \exp(-2\gamma)
  \int_0^{\infty} r \exp(-r^2/2)\, a_2(r)\, dr,
\end{equation}
where
\begin{equation}\label{cosineInt}
a_2(r)
= \frac{1}{2\pi}
  \int_0^{2\pi} (\cos p\theta)\,
    \exp\{2\sqrt{\gamma}\, r\cos\theta\}\, d\theta.
\end{equation}

From Appendix 1 of \citet{mardiajupp2000},
\begin{equation}\label{a2}
a_2(r) = I_p(2\sqrt{\gamma}\, r).
\end{equation}

Using \citet{gradshteyn1980}, formula 6.631(7), p.~706,
\begin{equation}\label{Bessel}
\int_0^{\infty} r \exp(-r^2/2)\,
I_p(2\sqrt{\gamma}\, r)\, dr
= \sqrt{\frac{\pi\gamma}{2}}\, \exp(\gamma)
  \{ I_{(p-1)/2}(\gamma) + I_{(p+1)/2}(\gamma) \}.
\end{equation}

Substituting (\ref{a2}) and (\ref{Bessel}) into (\ref{Ncosine}) yields (\ref{eq:cosMoment}).

\section{Appendix 3: Assessment of Approx 1 and Approx 2}\label{Sec: Appendix3Approx12}
\label{SuppSME}
\begin{figure}[!htb]
\centering
\includegraphics[width=.9\textwidth]{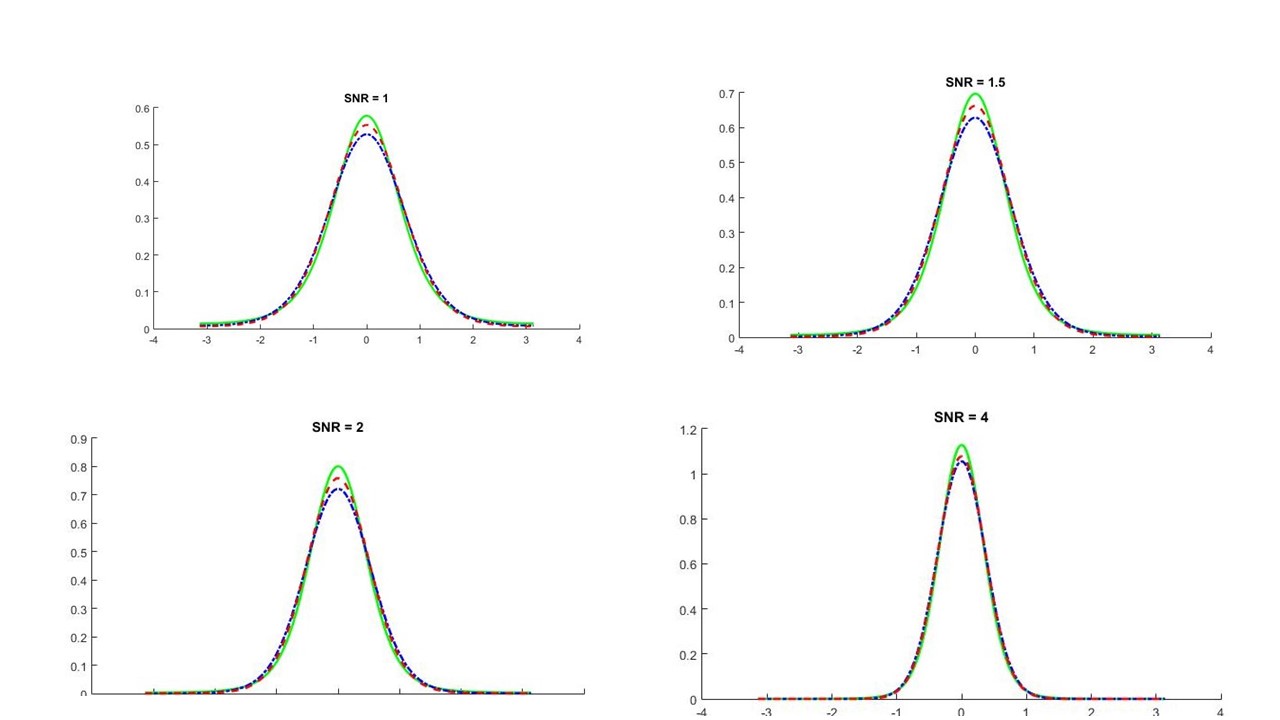}
\caption{Comparison of PIN density (solid green) with von Mises Approx 1 (dashed blue) and Approx 2 (dashed red) for some selected SNR values.}
\label{FigApproxNew}
\end{figure}

We now give some  more comparisons of Approx 1 and Approx 2 of Section \ref{Sec: Pop}.
\begin{table}[ht]
\centering
\caption{Von Mises approximations to $\mathrm{PIN}(\gamma)$; for given $\gamma$, $\kappa_1$ from Approx 1 and $\kappa_2$ from  Approx 2.}
\label{TwoAppr}
\small
\setlength{\tabcolsep}{6pt}
\begin{tabular}{l c c}
\hline
$\gamma$ & $\kappa_1$ & $\kappa_2$ \\
\hline
0.05  & 0.5686  & 0.5746  \\
0.25  & 1.3513  & 1.4161  \\
0.50  & 2.0786  & 2.2473  \\
0.75  & 2.7936  & 3.0642  \\
1.00  & 3.5628  & 3.9059  \\
2.00  & 7.2644  & 7.5655  \\
2.50  & 9.2872  & 9.5093  \\
3.75  & 14.3748 & 14.4765 \\
5.00  & 19.4204 & 19.4790 \\
\hline
\end{tabular}
\end{table}
Table \ref{TwoAppr} gives the concentration parameters \(\kappa\) for Approx 1  and Approx 2 ($\kappa_1$ and   $\kappa_2$ respectively)   to  $\gamma$ of the PIN distribution. From this table, we see that  $\kappa_2\ge\kappa_1$ for the values shown, so the difference \(\kappa_2-\kappa_1\) is nonnegative. Further, the maximum difference is approximately \(0.37\) near \(\gamma\approx 2.5\); for very small and very large \(\gamma\) the two approximations coincide to numerical precision.
\begin{figure}[!htb]
\centering
\includegraphics[width=\textwidth]{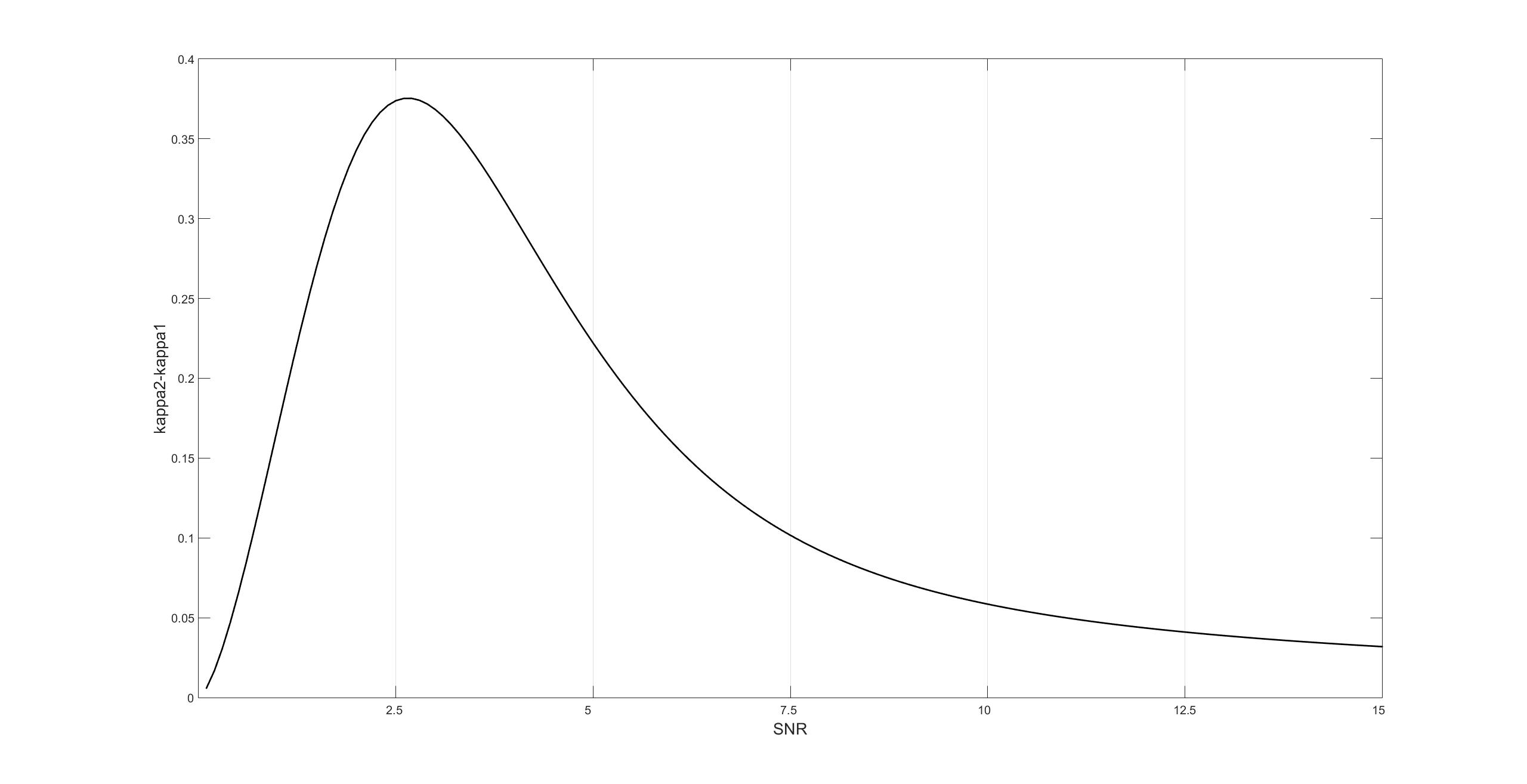}
\caption{Difference \(\kappa_2-\kappa_1\) as a function of SNR (where SNR=2$\gamma$).}
\label{fig:ApproxDiff}
\end{figure}
Figure \ref{fig:ApproxDiff} plots the difference between the two approximations; the largest discrepancy occurs in the moderate SNR range. Values are nonnegative as again confirming  \(\kappa_2\ge\kappa_1\). Figure \ref{FigApproxNew} compares the two von Mises approximations (Approx 1, Approx 2)  to the PIN density for  SNR values 1, 1.5, 2 and 4 where one expects to see the larger differences. Approx 2 is slightly closer to the PIN in the intermediate SNR range (roughly \(0.75\)–\(2\)), likely because it incorporates the second trigonometric moment. 
\subsection*{Kullback–Leibler comparison}
We assessed the two approximations using Kullback–Leibler (KL) divergence from the PIN density. The KL differences between Approx 1 and Approx 2 are negligible: the largest observed difference is about \(0.003\) at \(\gamma=0.75\), with Approx 1 being marginally better. Thus we note that:
\begin{itemize}
  \item Approx 1 is a well‑established choice and performs slightly better in some tail regions.
  \item Approx 2 is algebraically simpler and accounts for higher trigonometric moments, improving fit in the moderate SNR range.
  \item For practical purposes (e.g., thresholding \(\bar{R}\) in applications), the two approximations give nearly identical results; differences are most relevant only in the extreme tails.
\end{itemize}

\end{document}